\documentclass[10pt,conference]{IEEEtran}
\usepackage{times}
\usepackage{cite}
\usepackage{epsfig}
\usepackage{graphicx}
\usepackage{array}
\usepackage{amssymb}
\usepackage{mathrsfs}
\usepackage[dvipsnames]{xcolor}
\usepackage[colorlinks]{hyperref}
\newtheorem{Propo}{Proposition}[section]

\begin{document}
\pagenumbering{gobble}
\title{\textbf{\Large Near Capacity Signaling over Fading Channels using
Coherent Turbo Coded OFDM and Massive MIMO}}
\author{\IEEEauthorblockN{~\\[-0.4ex]\large K. Vasudevan\\[0.3ex]\normalsize}

\IEEEauthorblockA{Dept. of EE\\
                  IIT Kanpur\\
                  India\\
Email: {\tt vasu@iitk.ac.in}}
}
\maketitle
\begin{abstract}
\boldmath
The minimum average signal-to-noise ratio (SNR) per bit required for error-free
transmission over a fading channel is derived, and is shown to be equal to
that of the additive white Gaussian noise (AWGN)
channel, which is $-1.6$ dB. Discrete-time algorithms are presented for
timing and carrier synchronization, as well as channel estimation, for
turbo coded multiple input multiple output (MIMO) orthogonal frequency
division multiplexed (OFDM) systems. Simulation results show that it is
possible to achieve a bit error rate of $10^{-5}$ at an average SNR per bit
of 5.5 dB, using two transmit and two receive antennas. We then propose a
near-capacity signaling method in which each transmit antenna
uses a different carrier frequency. Using the near-capacity approach,
we show that it is possible to achieve a BER of $2\times 10^{-5}$ at an
average SNR per bit of just 2.5 dB, with
one receive antenna for each transmit antenna. When the number of
receive antennas for each transmit antenna is increased to 128, then a
BER of $2\times 10^{-5}$ is attained at an average SNR per bit of 1.25 dB.
In all cases, the number of
transmit antennas is two and the spectral efficiency is 1 bit/transmission or
1 bit/sec/Hz.
In other words, each transmit antenna sends 0.5 bit/transmission. It is
possible to obtain higher spectral efficiency by increasing the number of
transmit antennas, with no loss in BER performance, as long as each transmit
antenna uses a different carrier frequency. The transmitted signal spectrum
for the near-capacity approach can be restricted by pulse-shaping.
In all the simulations, a four-state turbo code is used. The corresponding
turbo decoder uses eight iterations.
The algorithms can be implemented on programmable hardware and there is a
large scope for parallel processing.
\end{abstract}
\begin{IEEEkeywords}
Channel capacity; coherent detection; frequency selective Rayleigh fading
channel; massive multiple input multiple output (MIMO);
orthogonal frequency division multiplexing (OFDM); spectral efficiency;
turbo codes.
\end{IEEEkeywords}
\IEEEpeerreviewmaketitle
\bstctlcite{IEEEexample:BSTcontrol}
\section{Introduction}
\label{Sec:Intro}
We begin this article with an open question: what is the operating
signal-to-noise (SNR) per bit or $E_b/N_0$ of the present day mobile phones
\cite{Vasu_ICWMC2016,Vasudevan2015,DBLP:journals/corr/Vasudevan15a}?
The mobile phones indicate a
typical received signal strength of $-100$ dBm ($10^{-10}$ mW), however this
is not the SNR per bit.

The above question assumes significance since future wireless communications,
also called the 5th generation or 5G
\cite{6824752,7067426,7399671,7414384}, is supposed to involve not only
billions of people,
but also smart machines and devices, e.g., driverless cars, remotely
controlled washing machines, refrigerators, microwave ovens, robotic
surgeries in health care and so on.
Thus, we have to deal with an internet of things (IoT), which involves
device-to-human, human-to-device and device-to-device communications.
Due to the large number of devices involved, it becomes
imperative that each device operates at the minimum possible average SNR per
bit required for error-free communication.

Depending on the application, there are different requirements on the
communication system. Critical applications like driverless cars and
robotic surgeries require low to medium bit rates, e.g., 0.1 -- 10 Mbps and low
latency (the time taken to process the received information and send a
response back to the transmitter) of the order of a fraction of a millisecond.
Some applications like watching movies on a mobile phone require high bit
rates, e.g., 10 -- 1000 Mbps for high density and ultra high density (4k) video
and can tolerate high latency, of the order of a fraction of a second. Whatever
the application, the 5G wireless communication systems are expected to share
some common features like having a large number of transmit and receive
antennas also called massive multiple input multiple output (MIMO)
\cite{6375940,6415388,7294693,7601145} and the use of millimeter wave carrier
frequencies ($>$ 100 GHz) \cite{5783993,6387266,6515173,6882985,7574380},
to accommodate large bit-rates ($>$ 1 Gbps) and large number of users. In this
paper we deal with the physical layer of wireless systems that are also
applicable to 5G. The main topics addressed in this work are timing and
carrier synchronization, channel estimation, turbo codes and orthogonal
frequency division multiplexing (OFDM). Recall that OFDM converts a
frequency selective channel into a flat channel \cite{Vasu_Book10,Vasu_Book16}.

Channel characteristics in the THz frequency range and at 17 GHz for 5G indoor
wireless systems is studied in \cite{7511280,7725441}.
Channel estimation for massive MIMO assuming spatial correlation between
the receive antennas is considered in \cite{7752722,7760363}. In
\cite{7571183}, a MIMO channel estimator and beamformer is described.
Uplink channel estimation using compressive sensing for millimeter wave,
multiuser MIMO systems is considered in \cite{7727995,7752516}.

Waveform design for spectral containment of the transmitted signal, is an
important aspect of wireless telecommunications, especially in the uplink,
where many users access a base station. We require that the signal from one
user does not interfere with the other user. This issue is addressed in
\cite{5753092,7022995,7469313,7509396,7510757,7564682,7566627,7744813,7744816,
7744817,7763353,7785158}.
Error control coding for 5G is discussed in \cite{7593095,7565482}.
References to carrier and timing synchronization
in OFDM can be found in
\cite{6663392,Vasudevan2015,DBLP:journals/corr/Vasudevan15a}.

The capacity of single-user MIMO systems
under different assumptions about the channel impulse response (also
called the channel state information or CSI) and the statistics of the
channel impulse response (also called channel distribution information or CDI)
is discussed in \cite{1203154}. The capacity of MIMO Rayleigh fading
channels in the presence of interference and receive correlation is
discussed in \cite{4907090}. The low SNR capacity of MIMO fading channels
with imperfect channel state information is presented in \cite{6809207}.

The main contribution of this paper is to
develop discrete-time algorithms for coherently detecting
multiple input, multiple output (MIMO),
orthogonal frequency division multiplexed (OFDM) signals, transmitted
over frequency selective Rayleigh fading channels. Carrier frequency offset and
additive white Gaussian noise (AWGN) are the other impairments considered in
this work. The minimum SNR per bit required for error-free transmission
over frequency selective MIMO fading channels is derived.
Finally we demonstrate how we can approach close to the channel capacity.

To the best of our knowledge, other than the work in \cite{6663392},
which deals with turbo coded single input single output (SISO) OFDM, and
\cite{Vasudevan2015,DBLP:journals/corr/Vasudevan15a}, which deal with turbo
coded single input multiple output (SIMO) OFDM,
discrete-time algorithms for the coherent
detection of turbo coded MIMO OFDM systems have not been discussed earlier
in the literature. Coherent detectors for AWGN channels is discussed in
\cite{Vasu_SIVP10,DBLP:journals/corr/Vasudevan15}.
Simulations results for a $2\times 2$
turbo coded MIMO OFDM system indicate that a BER of $10^{-5}$, is obtained
at an average SNR per bit of just 5.5 dB, which is a 2.5 dB improvement over the
performance given in \cite{Vasudevan2015}. If each transmit antenna
transmits at a different carrier frequency, then we show that it is possible
to achieve a BER of $2\times 10^{-5}$ at an average SNR per bit of just 2.5 dB,
with one receive antenna for each transmit antenna. When the number of
receive antennas for each transmit antenna is increased to 128, then a
BER of $2\times 10^{-5}$ is obtained at an average SNR per bit of 1.25 dB.
In all cases, the number of
transmit antennas is two and the spectral efficiency is 1 bit/transmission
or 1 bit/sec/Hz.
In other words, each transmit antenna sends 0.5 bit/transmission. It is
possible to obtain higher spectral efficiency by increasing the number of
transmit antennas, with no loss in BER performance, as long as each transmit
antenna uses a different carrier frequency. It is possible to band limit
the transmitted signal using pulse shaping.
In all the simulations, a four-state turbo code is used. The corresponding
turbo decoder uses eight iterations.

This paper is organized as follows. Section~\ref{Sec:Sys_Model} presents the
system model. The discrete-time algorithms and simulation results for the
coherent receiver are given in Section~\ref{Sec:Receiver}. Near-capacity
signaling is presented in Section~\ref{Sec:Near_Capacity}. Finally,
Section~\ref{Sec:Conclusions} concludes the paper.

\section{System Model}
\label{Sec:Sys_Model}
\begin{figure}[tbh]
\centering
\input{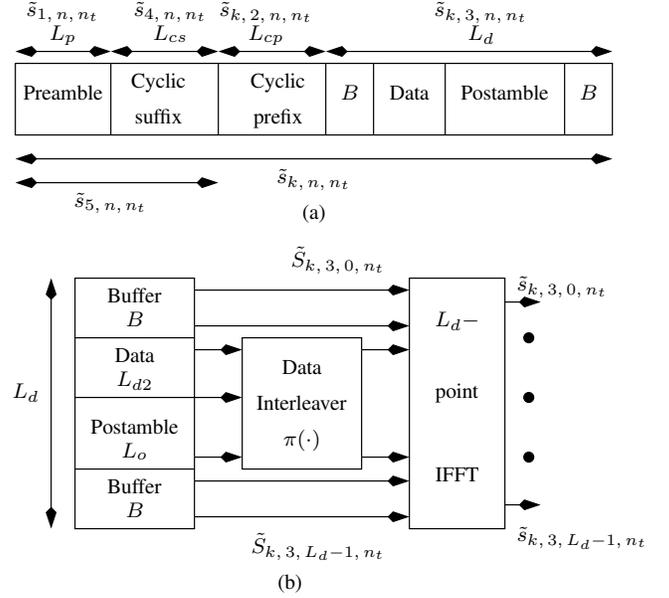}
\caption{The frame structure in the time domain.}
\label{Fig:Enh_Frame1}
\end{figure}

\begin{figure*}[tbh]
\centering
\input{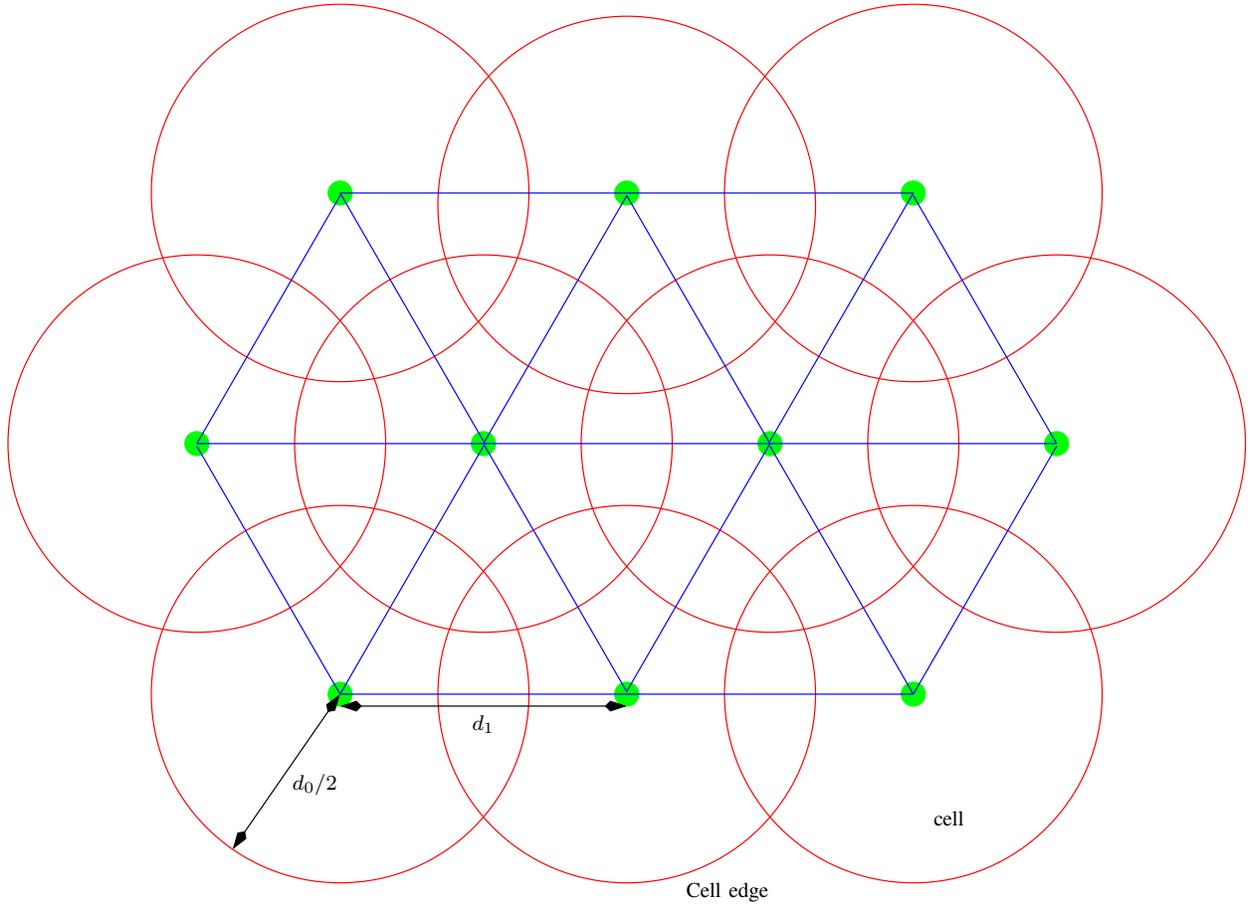}
\caption{Arrangement of cells and base stations.}
\label{Fig:System1}
\end{figure*}

We assume a MIMO-OFDM system with $N_t$ transmit and $N_r$ receive antennas,
with QPSK modulation.
The data from each transmit antenna is organized into frames, as shown in
Figure~\ref{Fig:Enh_Frame1}(a), similar to
\cite{6663392,Vasudevan2015,DBLP:journals/corr/Vasudevan15a}. Note the
presence of the cyclic suffix, whose purpose will be explained later.
In Figure~\ref{Fig:Enh_Frame1}(b), we observe that only the data and
postamble QPSK symbols are interleaved. The buffer QPSK symbols ($B$) are
sent to the IFFT without interleaving.
In Figure~\ref{Fig:Enh_Frame1}, the subscript $k$ refers to the $k^{th}$
frame, $n$ denotes the time index in a frame and $1 \le n_t \le N_t$ is the
index to the transmit antenna. The total length of the frame is
\begin{eqnarray}
\label{Eq:Pap9_Eq1}
L = L_p + L_{cs} + L_{cp} + L_d.
\end{eqnarray}
Let us assume a channel span equal to $L_h$. The channel span assumed by the
receiver is \cite{6663392,DBLP:journals/corr/Vasudevan15a}
\begin{eqnarray}
\label{Eq:Pap9_Eq1_0}
L_{hr}=2L_h-1
\end{eqnarray}
Note that $L_h$ depends on the delay spread
of the channel, and is measured in terms of the number of symbols.
Recall that, the delay spread is a measure of the time difference between the
arrival of the first and the last multipath signal, as seen by the receiver.
Typically
\begin{eqnarray}
\label{Eq:Pap9_Eq1_1}
L_h = d_0/(cT_s)
\end{eqnarray}
where $d_0$ is the distance between the longest and shortest
multipath, $c$ is the velocity of light and $T_s$ is the symbol duration
which is equal to the sample spacing of $\tilde s_{k,\, n,\, n_t}$ in
Figure~\ref{Fig:Enh_Frame1}(a).
We have assumed a situation where the mobile is close to the base station
and the longest path is reflected from the cell edge, which is
approximately equal to the cell diameter $d_0$, as shown in
Figure~\ref{Fig:System1}.
The base stations, depicted by green dots, are interconnected by a high
data-rate backhaul, shown by the blue lines. The cell edge is given by the
red circles. Note that $d_1<d_0$. In order to obtain
symmetry, the backhaul forms an equilateral triangle of length
$d_1$. The base station is at the center of each cell, whose
diameter is $d_0$.
For $L_h=10$, $1/T_s=10^7$ bauds and $c=3\times 10^8$ meters per sec, we get
$d_0=300$ meters.
Similarly with $L_h=10$ and $1/T_s=10^8$ bauds we obtain $d_0=30$ meters. In
other words, as the baud rate increases, the cell size needs to decrease,
and consequently the transmit power decreases, for the same channel span $L_h$.
The length of the cyclic prefix and suffix is \cite{Vasu_Book10}:
\begin{eqnarray}
\label{Eq:Pap9_Eq2}
L_{cp} = L_{cs} = L_{hr} - 1.
\end{eqnarray}

Throughout the manuscript, we use tilde  to denote complex quantities.
However, complex QPSK symbols will be denoted without a tilde, e.g.,
$S_{1,\, n,\, n_t}$. Boldface letters denote vectors or matrices.
The channel coefficients $\tilde h_{k,\, n,\, n_r,\, n_t}$ associated with
the receive antenna $n_r$ ($1 \le n_r \le N_r$) and transmit antenna $n_t$
($1 \le n_t \le N_t$) for the $k^{th}$ frame are
$\mathscr{CN}(0,\, 2\sigma^2_f)$ and independent over time $n$, that is:
\begin{eqnarray}
\label{Eq:Pap9_Eq3}
\frac{1}{2}
 E
\left[
\tilde h_{k,\, n,\, n_r,\, n_t}
\tilde h_{k,\, n-m,\, n_r,\, n_t}^*
\right] = \sigma^2_f \delta_K(m)
\end{eqnarray}
where ``*'' denotes complex conjugate and $\delta_K(\cdot)$ is the
Kronecker delta function. This implies a uniform power delay profile. Note
that even though an exponential power delay profile is more realistic,
we have used a uniform power delay profile, since it is expected to give the
worst-case BER performance, as all the multipath components have the same
power. The channel is assumed to be quasi-static, that is
$\tilde h_{k,\, n,\, n_r,\, n_t}$
is time-invariant over one frame and varies independently from frame-to-frame,
as given by:
\begin{eqnarray}
\label{Eq:Pap9_Eq4}
\frac{1}{2}
 E
\left[
\tilde h_{k,\, n,\, n_r,\, n_t}
\tilde h_{j,\, n,\, n_r,\, n_t}^*
\right] = \sigma^2_f \delta_K(k-j)
\end{eqnarray}
where $k$ and $j$ denote the frame indices.

The AWGN noise samples
$\tilde w_{k,\, n,\, n_r}$ for the $k^{th}$ frame at time $n$  and receive
antenna $n_r$ are $\mathscr{CN}(0,\, 2\sigma^2_w)$. The frequency offset
$\omega_k$ for the $k^{th}$ frame is uniformly distributed over
$[-0.04,\, 0.04]$ radian \cite{Minn03}.
We assume that  $\omega_k$ is fixed for a frame and varies randomly
from frame-to-frame.
\begin{figure}[tbh]
\centering
\input{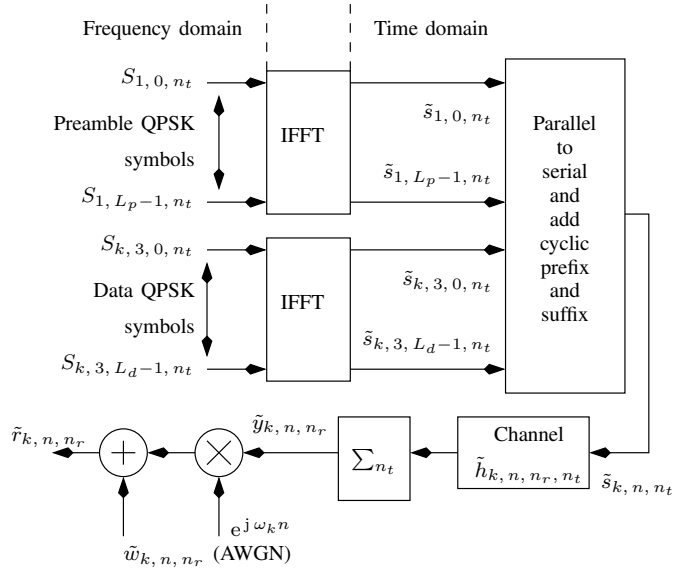}
\caption{Block diagram of the transmitter.}
\label{Fig:Frame}
\end{figure}

The block diagram of the transmitter is given in Figure~\ref{Fig:Frame}.
With reference to Figs.~\ref{Fig:Enh_Frame1}(a) and \ref{Fig:Frame}, note that:
\begin{eqnarray}
\label{Eq:Pap9_Eq4_1}
\tilde s_{1,\, n,\, n_t}
                  & = & \frac{1}{L_p}
                        \sum_{i=0}^{L_p-1}
                         S_{1,\, i,\, n_t}
                        \mathrm{e}^{\,\mathrm{j}\, 2\pi ni/L_p}
                        \nonumber  \\
                  &   & \mbox{ } \quad
                        \mbox{for $0\le n \le L_p-1$}      \nonumber  \\
\tilde s_{k,\, 3,\, n,\, n_t}
                  & = & \frac{1}{L_d}
                        \sum_{i=0}^{L_d-1}
                         S_{k,\, 3,\, i,\, n_t}
                        \mathrm{e}^{\,\mathrm{j}\, 2\pi ni/L_d}
                        \nonumber  \\
                  &   & \mbox{ } \quad
                        \mbox{for $0\le n \le L_d-1$}      \nonumber  \\
\tilde s_{k,\, 2,\, n,\, n_t}
                  & = & \tilde s_{k,\, 3,\, L_d-L_{cp}+n,\, n_t}
                        \nonumber  \\
                  &   & \mbox{ } \quad
                        \mbox{for $0 \le n \le L_{cp}-1$}  \nonumber  \\
\tilde s_{4,\, n,\, n_t}
                  & = & \tilde s_{1,\, n,\, n_t}
                        \nonumber  \\
                  &   & \mbox{ } \quad
                        \mbox{for $0 \le n \le L_{cs}-1$}  \nonumber  \\
\tilde s_{5,\, n,\, n_t}
                  & = & \tilde s_{1,\, n,\, n_t} +
                        \tilde s_{4,\, n-L_p,\, n_t}.
\end{eqnarray}
From (\ref{Eq:Pap9_Eq4_1}), it is clear that the preamble is independent of
the frame $k$. However, each transmit antenna has its own preamble, for the
purpose of synchronization and channel estimation at the receiver.

The preamble in the frequency domain, for each transmit antenna is generated
as follows. Let $\pi_p(i)$, for $0 \le i \le L_p -1$, denote the interleaver
map for the preamble. Let
\begin{eqnarray}
\label{Eq:Pap9_Eq4_1_0}
\mathbf{S}_r =
\left[
\begin{array}{ccc}
S_{r,\, 0} & \ldots & S_{r,\, L_p - 1}
\end{array}
\right]^T_{L_p\times 1}
\end{eqnarray}
denote a random vector of QPSK symbols. The preamble vector for the
transmit antenna $n_t$ is first initialized by
\begin{eqnarray}
\label{Eq:Pap9_Eq4_1_0_1}
\mathbf{S}_{1,\, n_t}
& = &
\left[
\begin{array}{ccc}
S_{1,\, 0,\, n_t} & \ldots & S_{1,\, L_p - 1,\, n_t}
\end{array}
\right]^T_{L_p\times 1}     \nonumber  \\
& = & \mathbf{0}_{L_p\times 1}.
\end{eqnarray}
Next, we substitute
\begin{eqnarray}
\label{Eq:Pap9_Eq4_1_0_2}
\mathbf{S}_{1,\, \pi_p(i_4:i_5),\, n_t} = \mathbf{S}_r(i_4:i_5).
\end{eqnarray}
where $i_4:i_5$ denotes the range of indices from $i_4$ to $i_5$, both
inclusive, and
\begin{eqnarray}
\label{Eq:Pap9_Eq4_1_0_3}
i_4 & = & (n_t - 1)L_p/N_t    \nonumber  \\
i_5 & = & i_4 + L_p/N_t -1.
\end{eqnarray}
Note that the preamble in the frequency domain for each transmit antenna
has only $L_p/N_t$ non-zero elements, the rest of the elements are zero.
Moreover, due to $\pi_p(\cdot)$, the $L_p/N_t$ non-zero elements are randomly
interspersed over the $L_p$ subcarriers in the frequency domain, for each
transmit antenna.

By virtue of the preamble construction in (\ref{Eq:Pap9_Eq4_1_0_1}),
(\ref{Eq:Pap9_Eq4_1_0_2}) and (\ref{Eq:Pap9_Eq4_1_0_3}), the preambles in the
frequency and time domains corresponding to transmit antennas $n_t$ and $m_t$
satisfy the relation (using Parseval's energy theorem):
\begin{eqnarray}
\label{Eq:Pap9_Eq4_1_1}
S_{1,\, i,\, n_t}
S_{1,\, i,\, m_t}^*        & = & (2N_tL_p/L_d)
                                 \delta_K(n_t - m_t)            \nonumber  \\
                           &   & \mbox{ } \quad
                                 \mbox{for $0 \le i \le L_p-1$} \nonumber  \\
\Rightarrow
\tilde s_{1,\,  n,\, n_t} \odot_{L_p}
\tilde s_{1,\, -n,\, m_t}^*
                           & = & \left
                                 \{
                                 \begin{array}{l}
                                  0
                                 \quad
                                 \mbox{for $n_t \ne m_t$,}\\
                                 \mbox{ }
                                 \quad
                                 \mbox{$0\le n\le L_p - 1$}\\
                                 (2L_p/L_d) \delta_K(n)\\
                                 \mbox{ }
                                 \quad
                                 \mbox{for $n_t = m_t$}
                                 \end{array}
                                 \right.
\end{eqnarray}
where ``$\odot_{L_p}$'' denotes the $L_p$-point circular convolution. In other
words, the preambles corresponding to distinct transmit antennas are
orthogonal over $L_p$ samples. Moreover, the autocorrelation of the preambles
in frequency and time domain, can be approximated by a weighted Kronecker
delta function (this condition is usually satisfied by random
sequences having zero-mean; the approximation gets better as $L_p$ increases).

We assume $S_{k,\, 3,\, i,\, n_t}\in \{\pm 1 \pm \mathrm{j}\}$. Since we
require:
\begin{eqnarray}
\label{Eq:Pap9_Eq4_2}
 E
\left[
\left|
\tilde s_{1,\, n,\, n_t}
\right|^2
\right] & = &
 E
\left[
\left|
\tilde s_{k,\, 3,\, n,\, n_t}
\right|^2
\right] = 2/L_d \stackrel{\Delta}{=} \sigma^2_s
\end{eqnarray}
we must have $S_{1,\, i,\, n_t}\in \sqrt{L_p N_t/L_d}\,(\pm 1 \pm \mathrm{j})$.
In other words, the average power of the preamble part must be equal to the
average power of the data part, in the time domain.

Due to the presence of the cyclic suffix
in Figure~\ref{Fig:Enh_Frame1} and (\ref{Eq:Pap9_Eq4_1}), and due to
(\ref{Eq:Pap9_Eq4_1_1}), we have
\begin{eqnarray}
\label{Eq:Pap9_Eq4_1_2}
\lefteqn{
\tilde s_{5,\, n,\, n_t}
\star
\tilde s_{1,\, L_p-1-n,\, m_t}^*}                       \nonumber  \\
& = &
\left
\{
\begin{array}{l}
 0       \quad   \mbox{for $L_p-1\le n\le L_p+L_{hr}-2$,}\\
\mbox{ } \quad   \mbox{$n_t \ne m_t$}\\
(2L_p/L_d)
\delta_K(n-L_p+1)\\
\mbox{ } \quad    \mbox{for $n_t = m_t$}
\end{array}
\right.
\end{eqnarray}
where ``$\star$'' denotes linear convolution.

The signal for the $k^{th}$ frame and receive antenna $n_r$ can be written
as (for $0 \le n \le L+L_h-2$):
\begin{eqnarray}
\label{Eq:Pap9_Eq5}
\tilde r_{k,\, n,\, n_r}
                  & = & \sum_{n_t=1}^{N_t}
                        \left(
                        \tilde s_{k,\, n,\, n_t}
                        \star
                        \tilde h_{k,\, n,\, n_r,\, n_t}
                        \right)\,
                        \mathrm{e}^{\,\mathrm{j}\,\omega_k n} +
                        \tilde w_{k,\, n,\, n_r}                 \nonumber  \\
                  & = & \tilde y_{k,\, n,\, n_r}
                        \mathrm{e}^{\,\mathrm{j}\,\omega_k n} +
                        \tilde w_{k,\, n,\, n_r}
\end{eqnarray}
where $\tilde s_{k,\, n,\, n_t}$ is depicted in Figure~\ref{Fig:Enh_Frame1}(a)
and
\begin{eqnarray}
\label{Eq:Pap9_Eq5_1}
\tilde y_{k,\, n,\, n_r} = \sum_{n_t=1}^{N_t}
                           \tilde s_{k,\, n,\, n_t}
                           \star
                           \tilde h_{k,\, n,\, n_r,\, n_t}.
\end{eqnarray}
Note that any random carrier phase can be absorbed in the
channel impulse response.

The uplink and downlink transmissions between the mobiles and base station
could be carried out using time division duplex (TDD) or frequency division
duplex (FDD). Time division (TDMA), frequency division (FDMA), code division
(CDMA), orthogonal frequency division (OFDMA), for downlink transmissions
and filterbank multicarrier (FBMC), for uplink transmissions \cite{5753092},
are the possible choices for multiple access (MA) techniques.

\section{Receiver}
\label{Sec:Receiver}
In this section, we discuss the discrete-time receiver algorithms.
\subsection{Start of Frame (SoF) and Coarse Frequency Offset Estimate}
\label{SSec:SoF}
The start of frame (SoF)
detection and coarse frequency offset estimation is performed for each receive
antenna $1\le n_r\le N_r$ and transmit antenna $1\le n_t\le N_t$,
as given by the following rule (similar to (22) in \cite{6663392} and (24) in
\cite{DBLP:journals/corr/Vasudevan15a}): choose that value of $m$
and $\nu_k$ which maximizes
\begin{eqnarray}
\label{Eq:Pap9_Eq16}
\left|
\left(
\tilde r_{k,\, m,\, n_r}
\,
\mathrm{e}^{-\mathrm{j}\,\nu_k m}
\right)
\star
\tilde s_{1,\, L_p-1-m,\, n_t}^*
\right|.
\end{eqnarray}
Let $\hat m_k(\cdot)$ denote the time instant and $\hat\nu_k(\cdot)$ denote
the coarse estimate of the frequency offset (both of which are functions
of $n_r$ and $n_t$), at which the maximum in (\ref{Eq:Pap9_Eq16}) is obtained.
Note that (\ref{Eq:Pap9_Eq16}) is a two-dimensional search over $m$ and
$\nu_k$, which can be efficiently implemented in hardware, and there is a
large scope for parallel processing. In particular, the search over $\nu_k$
involves dividing the range of $\omega_k$ ($[-0.04,\, 0.04]]$ radians) into
$B_1$ frequency bins, and deciding in favour of that bin which maximizes
(\ref{Eq:Pap9_Eq16}). In our simulations, $B_1=64$
\cite{6663392,DBLP:journals/corr/Vasudevan15a}.

Note that in the absence
of noise and due to the properties given in (\ref{Eq:Pap9_Eq4_1_2})
\begin{eqnarray}
\label{Eq:Pap9_Eq17}
\hat m_k(n_r,\, n_t) = L_p - 1 + \mbox{argmax}_m
                                 \left|
                                 \tilde h_{k,\, m,\, n_r,\, n_t}
                                 \right|
\end{eqnarray}
where $\mbox{argmax}_m$ corresponds to the value of $m$ for which
$\left|\tilde h_{k,\, m,\, n_r,\, n_t}\right|$ is maximum. We also have
\begin{eqnarray}
\label{Eq:Pap9_Eq18}
L_p - 1 \le \hat m_k(n_r,\, n_t) \le L_p + L_h - 2.
\end{eqnarray}
\begin{figure*}[tbh]
\centering
\includegraphics[scale=0.6]{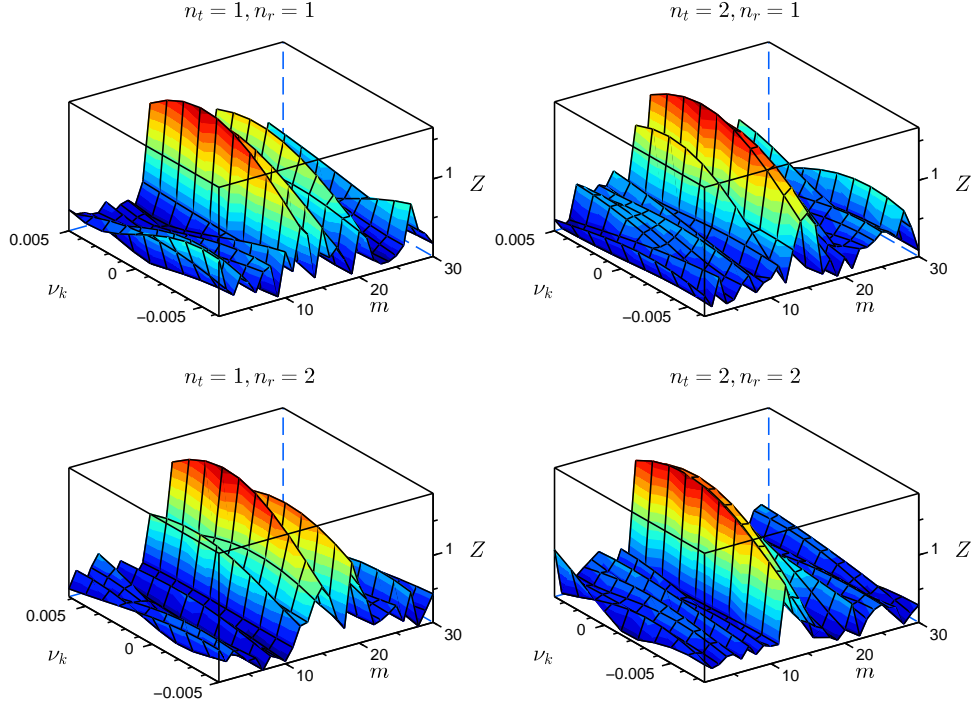}
\caption{SoF detection at 0 dB SNR per bit, $L_p=512$.}
\label{Fig:SoF_0db_Lp512}
\end{figure*}
\begin{figure*}[tbh]
\centering
\includegraphics[scale=0.6]{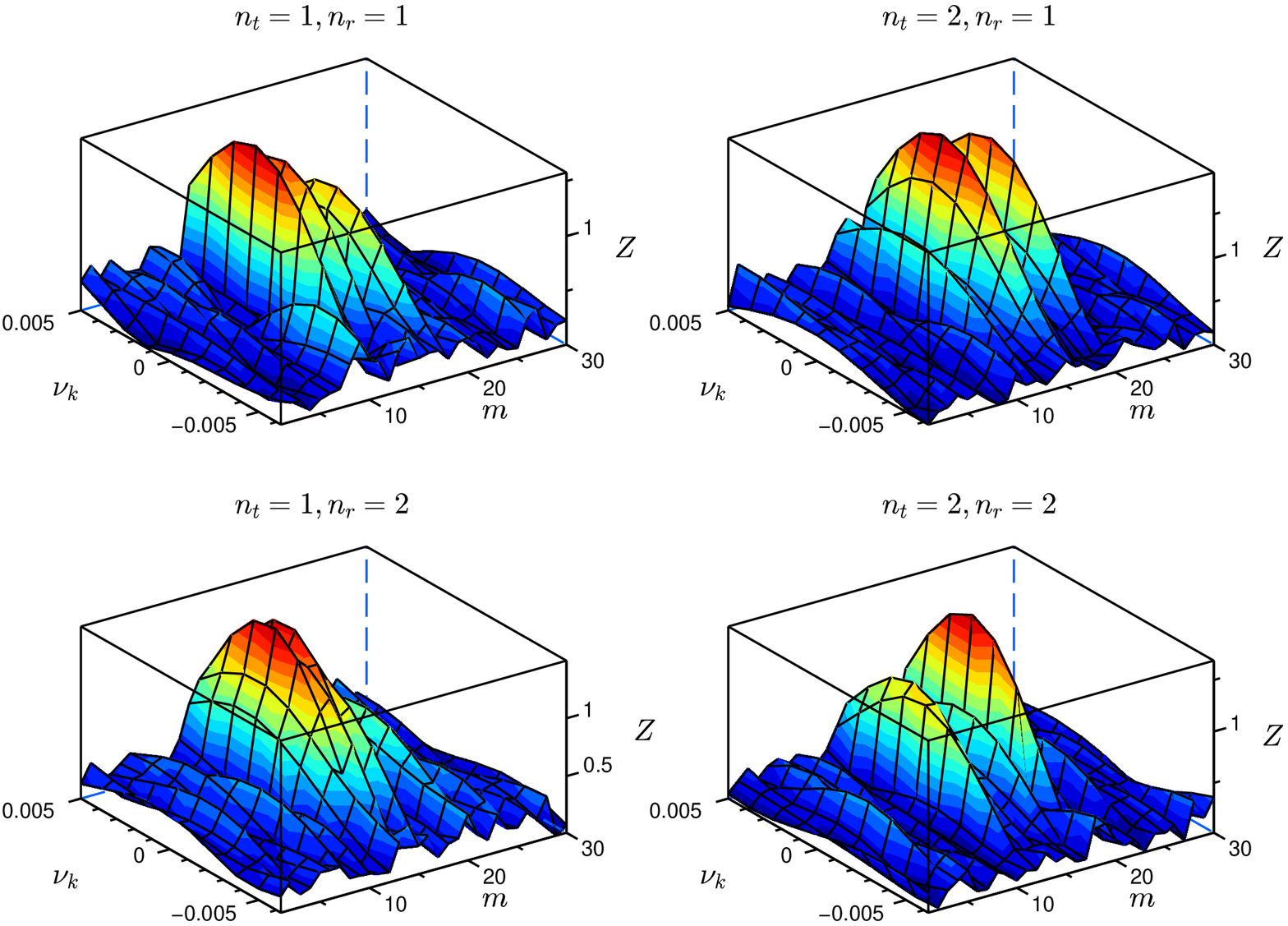}
\caption{SoF detection at 0 dB SNR per bit, $L_p=1024$.}
\label{Fig:SoF_0db_Lp1024}
\end{figure*}
\begin{figure*}[tbh]
\centering
\includegraphics[scale=0.6]{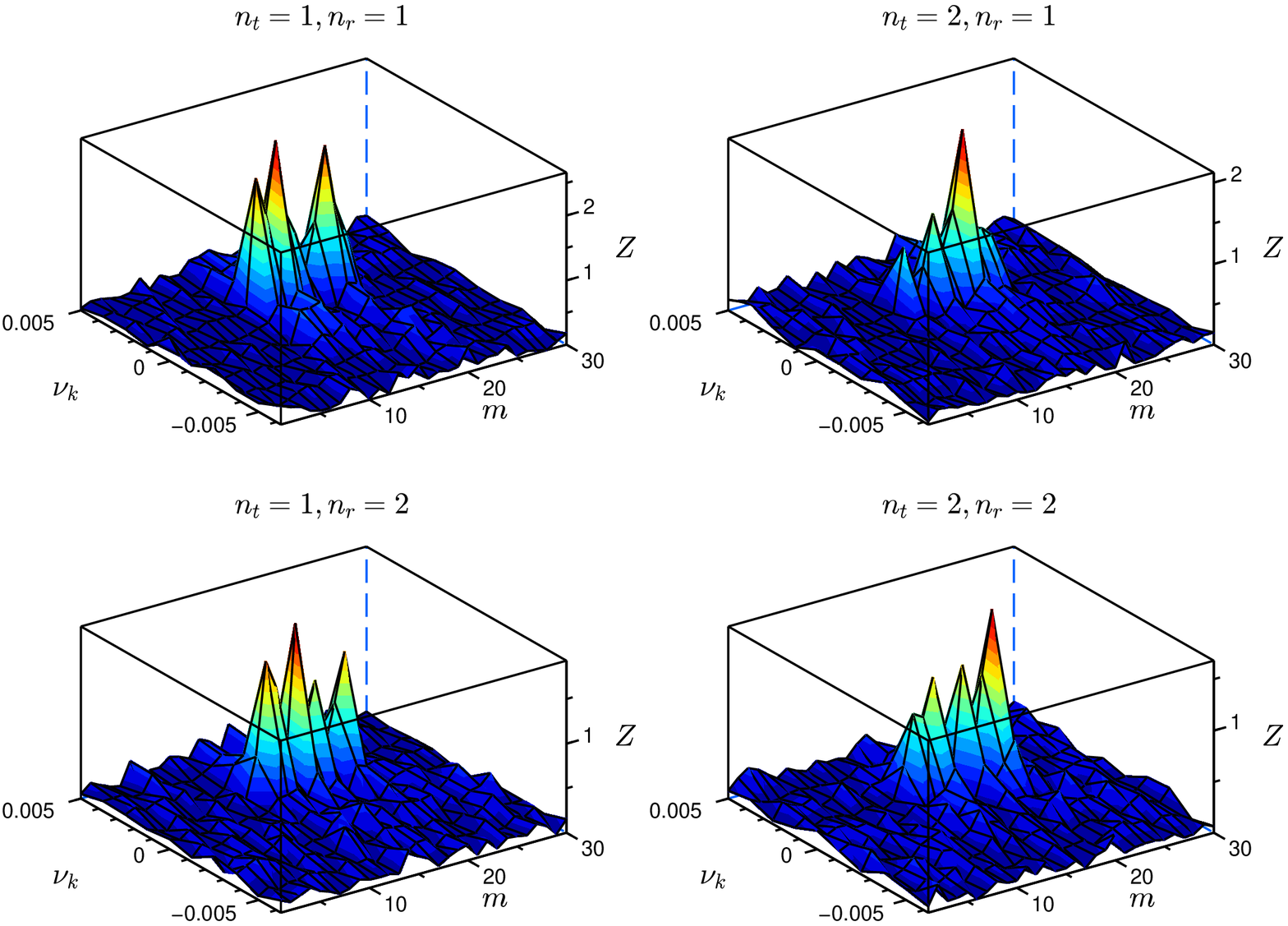}
\caption{SoF detection at 0 dB SNR per bit, $L_p=4096$.}
\label{Fig:SoF_0db_Lp4096}
\end{figure*}
If $\hat m_k(\cdot)$ lies outside the range in (\ref{Eq:Pap9_Eq18}),
the frame is declared as erased (not detected). This
implies that the peak in (\ref{Eq:Pap9_Eq16}) is due to noise, and not due
to the channel. The results for SoF detection at 0 dB SNR per bit for
$L_p=512,\, 1024,\, 4096$ are given in Figs.~\ref{Fig:SoF_0db_Lp512},
\ref{Fig:SoF_0db_Lp1024}, and \ref{Fig:SoF_0db_Lp4096}, respectively, for
$N_t=N_r=2$. The parameter $Z$ in the three figures denotes the correlation
magnitude given by (\ref{Eq:Pap9_Eq16}).

The average value of the coarse frequency offset estimate is
given by
\begin{eqnarray}
\label{Eq:Pap9_Eq19}
\hat\omega_k = \frac{\sum_{n_r=1}^{N_r}\sum_{n_t=1}^{N_t}
               \hat\nu_k(n_r,\, n_t)}
                    {N_r N_t}.
\end{eqnarray}
\subsection{Channel Estimation}
\label{SSec:Channel_Est}
We assume that the SoF has been estimated using (\ref{Eq:Pap9_Eq16}) with
outcome $m_{0,\, k}$ given by (assuming the condition in (\ref{Eq:Pap9_Eq18})
is satisfied for all $n_r$ and $n_t$):
\begin{eqnarray}
\label{Eq:Pap9_Eq19_1}
m_{0,\, k} = \hat m_k(1,\, 1) - L_p + 1
             \quad
             \mbox{$0 \le m_{0,\, k}\le L_h -1$}
\end{eqnarray}
and the frequency offset has been perfectly canceled
\cite{6663392,DBLP:journals/corr/Vasudevan15a}. Observe that any value of
$n_r$ and $n_t$ can be used in the computation of (\ref{Eq:Pap9_Eq19_1}).
We have taken $n_r=n_t=1$. Define
\begin{eqnarray}
\label{Eq:Pap9_Eq18_0}
m_{1,\, k} = m_{0,\, k} + L_h -1.
\end{eqnarray}
For the sake of notational simplicity, we drop the subscript $k$ in
$m_{1,\, k}$, and refer to it as $m_1$. The steady-state,
preamble part of the received signal for the $k^{th}$ frame and receive
antenna $n_r$ can be written as:
\begin{eqnarray}
\label{Eq:Pap9_Eq18_1}
\tilde\mathbf{r}_{k,\, m_1,\, n_r} =
\sum_{n_t=1}^{N_t}
\tilde\mathbf{s}_{5,\, n_t} \tilde\mathbf{h}_{k,\, n_r,\, n_t} +
\tilde\mathbf{w}_{k,\, m_1,\, n_r}
\end{eqnarray}
where
\begin{eqnarray}
\label{Eq:Pap9_Eq19_2}
\tilde\mathbf{r}_{k,\, m_1,\, n_r}
& = &
\left[
\begin{array}{ccc}
\tilde r_{k,\, m_1,\, n_r} & \ldots  & \tilde r_{k,\, m_1+L_p-1,\, n_r}
\end{array}
\right]^T                             \nonumber  \\
&   & \mbox{ }
      [L_p\times 1]
      \quad \mbox{vector}             \nonumber  \\
\tilde\mathbf{w}_{k,\, m_1,\, n_r}
& = &
\left[
\begin{array}{ccc}
\tilde w_{k,\, m_1,\, n_r} & \ldots  & \tilde w_{k,\, m_1+L_p-1,\, n_r}
\end{array}
\right]^T                             \nonumber  \\
&   & \mbox{ }
      [L_p\times 1]
      \quad \mbox{vector}             \nonumber  \\
\tilde\mathbf{h}_{k,\, n_r,\, n_t}
& = &
\left[
\begin{array}{ccc}
\tilde h_{k,\, 0,\, n_r,\, n_t}
                           & \ldots  &
\tilde h_{k,\, L_{hr}-1,\, n_r,\, n_t}
\end{array}
\right]^T                             \nonumber  \\
&   & \mbox{ }
      [L_{hr}\times 1]
      \quad \mbox{vector}             \nonumber  \\
\tilde\mathbf{s}_{5,\, n_t}
& = &
\left[
\begin{array}{ccc}
\tilde s_{5,\, L_{hr}-1,\, n_t}
                           & \ldots  & \tilde s_{5,\, 0,\, n_t}\\
\vdots                     & \ldots  & \vdots\\
\tilde s_{5,\, L_p+L_{hr}-2,\, n_t}
                           & \ldots  & \tilde s_{5,\, L_p-1,\, n_t}
\end{array}
\right]                               \nonumber  \\
&   & \mbox{ }
      [L_p\times L_{hr}]
      \quad \mbox{matrix}
\end{eqnarray}
where $L_{hr}$ is the channel length assumed by the receiver
(see (\ref{Eq:Pap9_Eq1_0})), $\tilde\mathbf{s}_{5,\, n_t}$ is the
channel estimation matrix and $\tilde\mathbf{r}_{k,\, m_1,\, n_r}$ is the
received signal vector {\it after\/} cancellation of the frequency offset.
Observe that $\tilde\mathbf{s}_{5,\, n_t}$ is
independent of $m_1$ and due to the relations in
(\ref{Eq:Pap9_Eq4_1_1}) and (\ref{Eq:Pap9_Eq4_1_2}), we have
\begin{eqnarray}
\label{Eq:Pap9_Eq19_3}
\tilde\mathbf{s}_{5,\, m_t}^H
\tilde\mathbf{s}_{5,\, n_t} =
\left
\{
\begin{array}{ll}
\mathbf{0}_{L_{hr}\times L_{hr}} & \mbox{for $n_t \ne m_t$}\\
(2L_p/L_d)
\mathbf{I}_{L_{hr}}              & \mbox{for $n_t = m_t$}
\end{array}
\right.
\end{eqnarray}
where $\mathbf{I}_{L_{hr}}$ is an $L_{hr}\times L_{hr}$ identity matrix and
$\mathbf{0}_{L_{hr}\times L_{hr}}$ is an $L_{hr}\times L_{hr}$ null matrix.
The statement of the ML channel estimation is as follows. Find
$\hat\mathbf{h}_{k,\, n_r,\, m_t}$ (the estimate of
$\tilde\mathbf{h}_{k,\, n_r,\, m_t}$) such that:
\begin{eqnarray}
\label{Eq:Pap9_Eq20}
\left(
\tilde\mathbf{r}_{k,\, m_1,\, n_r} -
\sum_{m_t=1}^{N_t}
\tilde\mathbf{s}_{5,\, m_t}
\hat\mathbf{h}_{k,\, n_r,\, m_t}
\right)^H      \nonumber  \\
\left(
\tilde\mathbf{r}_{k,\, m_1,\, n_r} -
\sum_{m_t=1}^{N_t}
\tilde\mathbf{s}_{5,\, m_t}
\hat\mathbf{h}_{k,\, n_r,\, m_t}
\right)
\end{eqnarray}
is minimized. Differentiating with respect to
$\hat\mathbf{h}_{k,\, n_r,\, m_t}^*$ and
setting the result to zero yields \cite{Haykin_Adapt_96,Vasu_Book10}:
\begin{eqnarray}
\label{Eq:Pap9_Eq21}
\hat\mathbf{h}_{k,\, n_r,\, m_t} =
                 \left(
                 \tilde\mathbf{s}_{5,\, m_t}^H
                 \tilde\mathbf{s}_{5,\, m_t}
                 \right)^{-1}
                 \tilde\mathbf{s}_{5,\, m_t}^H
                 \tilde\mathbf{r}_{k,\, m_1,\, n_r}.
\end{eqnarray}
Observe that when $m_{0,\, k}=L_h-1$ in (\ref{Eq:Pap9_Eq19_1}), and
noise is absent (see (29) in \cite{6663392} and (35) in
\cite{DBLP:journals/corr/Vasudevan15a}), we obtain:
\begin{eqnarray}
\label{Eq:Pap9_Eq22_1}
\lefteqn{
\hat\mathbf{h}_{k,\, n_r,\, m_t}} \hspace*{3in} \nonumber  \\
 =
\left[
\begin{array}{llllll}
\tilde h_{k,\, 0,\, n_r,\, m_t}     &
\ldots                              &
\tilde h_{k,\, L_h-1,\, n_r,\, m_t} &
0                                   &
\ldots                              &
0
\end{array}
\right]^T.
\end{eqnarray}
Similarly, when $m_{0,\, k}=0$ and in the absence of noise:
\begin{eqnarray}
\label{Eq:Pap9_Eq22_2}
\lefteqn{
\hat\mathbf{h}_{k,\, n_r,\, m_t}} \hspace*{3in} \nonumber  \\
 =
\left[
\begin{array}{llllll}
0                                   &
\ldots                              &
0                                   &
\tilde h_{k,\, 0,\, n_r,\, m_t}     &
\ldots                              &
\tilde h_{k,\, L_h-1,\, n_r,\, m_t}
\end{array}
\right]^T.
\end{eqnarray}
To see the effect of noise on the channel estimate in (\ref{Eq:Pap9_Eq21}),
consider
\begin{eqnarray}
\label{Eq:Pap9_Eq22}
\tilde\mathbf{u} =
                 \left(
                 \tilde\mathbf{s}_{5,\, m_t}^H
                 \tilde\mathbf{s}_{5,\, m_t}
                 \right)^{-1}
                 \tilde\mathbf{s}_{5,\, m_t}^H
                 \tilde\mathbf{w}_{k,\, m_1,\, n_r}.
\end{eqnarray}
It can be shown that
\begin{eqnarray}
\label{Eq:Pap9_Eq24}
 E
\left[
\tilde\mathbf{u}
\tilde\mathbf{u}^H
\right] = \frac{\sigma^2_w L_d}{L_p}
          \mathbf{I}_{L_{hr}}
          \stackrel{\Delta}{=}  2
          \sigma^2_u \mathbf{I}_{L_{hr}}.
\end{eqnarray}
Therefore, the variance of the ML channel estimate ($\sigma^2_u$) tends to
zero as $L_p\rightarrow \infty$ and $L_d$ is kept fixed. Conversely, when
$L_d$ is increased keeping $L_p$ fixed, there is noise enhancement
\cite{Vasudevan2015,DBLP:journals/corr/Vasudevan15a}.
\begin{figure*}[tbh]
\centering
\includegraphics[scale=0.6]{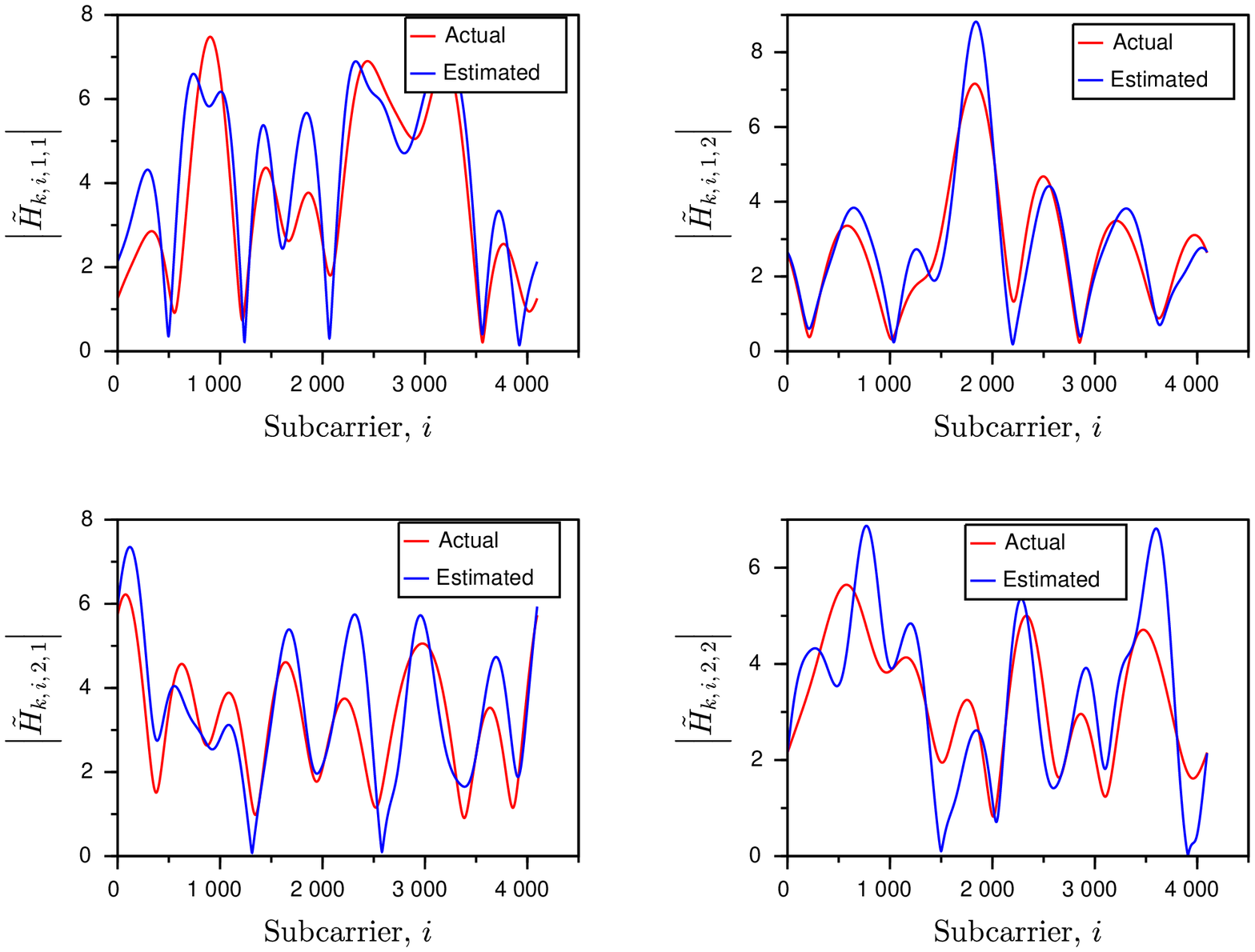}
\caption{Magnitude spectrum of estimated and actual channel, $L_p=512$.}
\label{Fig:DMT36_E_Chest_0DB_Lp512}
\end{figure*}
\begin{figure*}[tbh]
\centering
\includegraphics[scale=0.6]{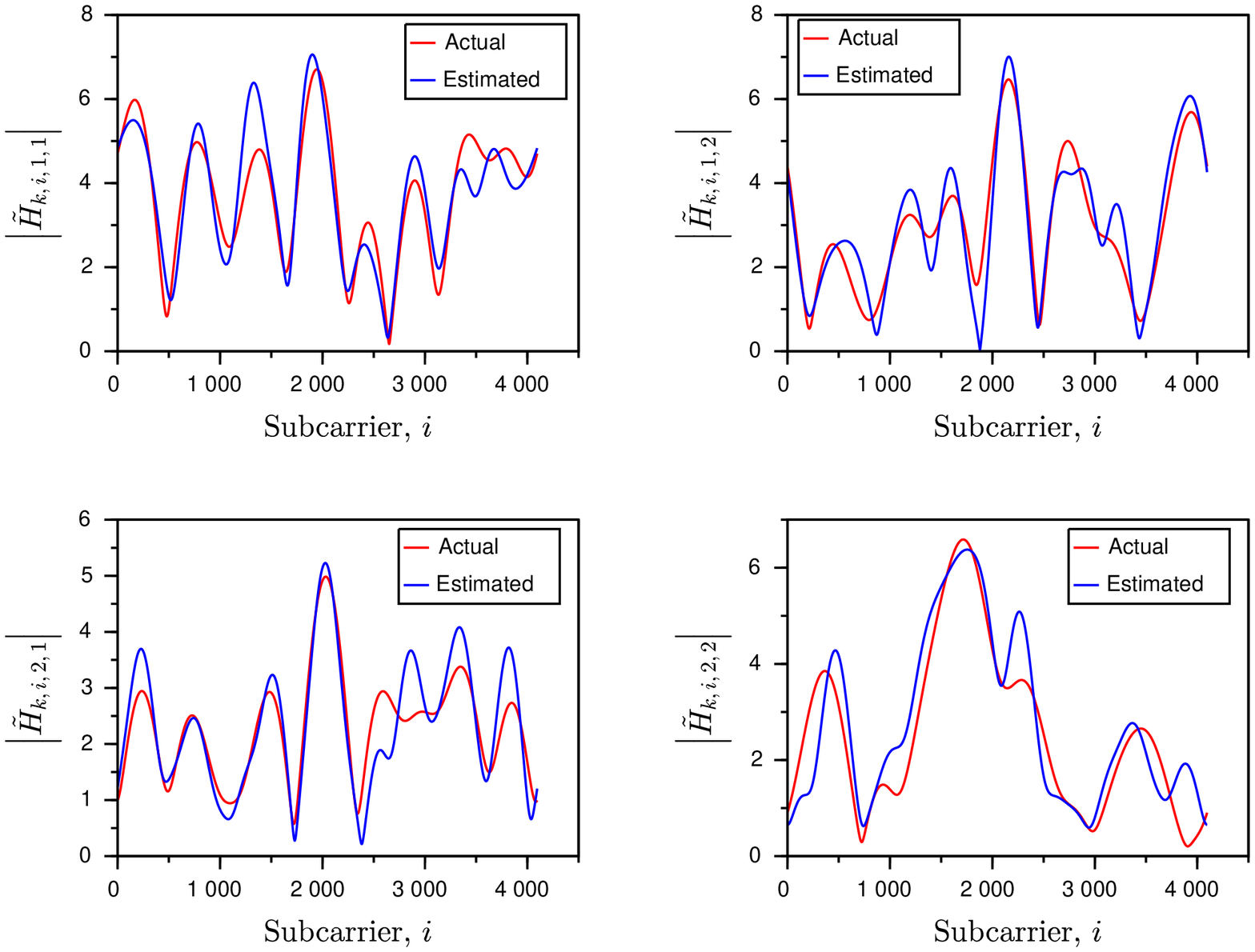}
\caption{Magnitude spectrum of estimated and actual channel, $L_p=1024$.}
\label{Fig:DMT36_E_Chest_0DB_Lp1024}
\end{figure*}
\begin{figure*}[tbh]
\centering
\includegraphics[scale=0.6]{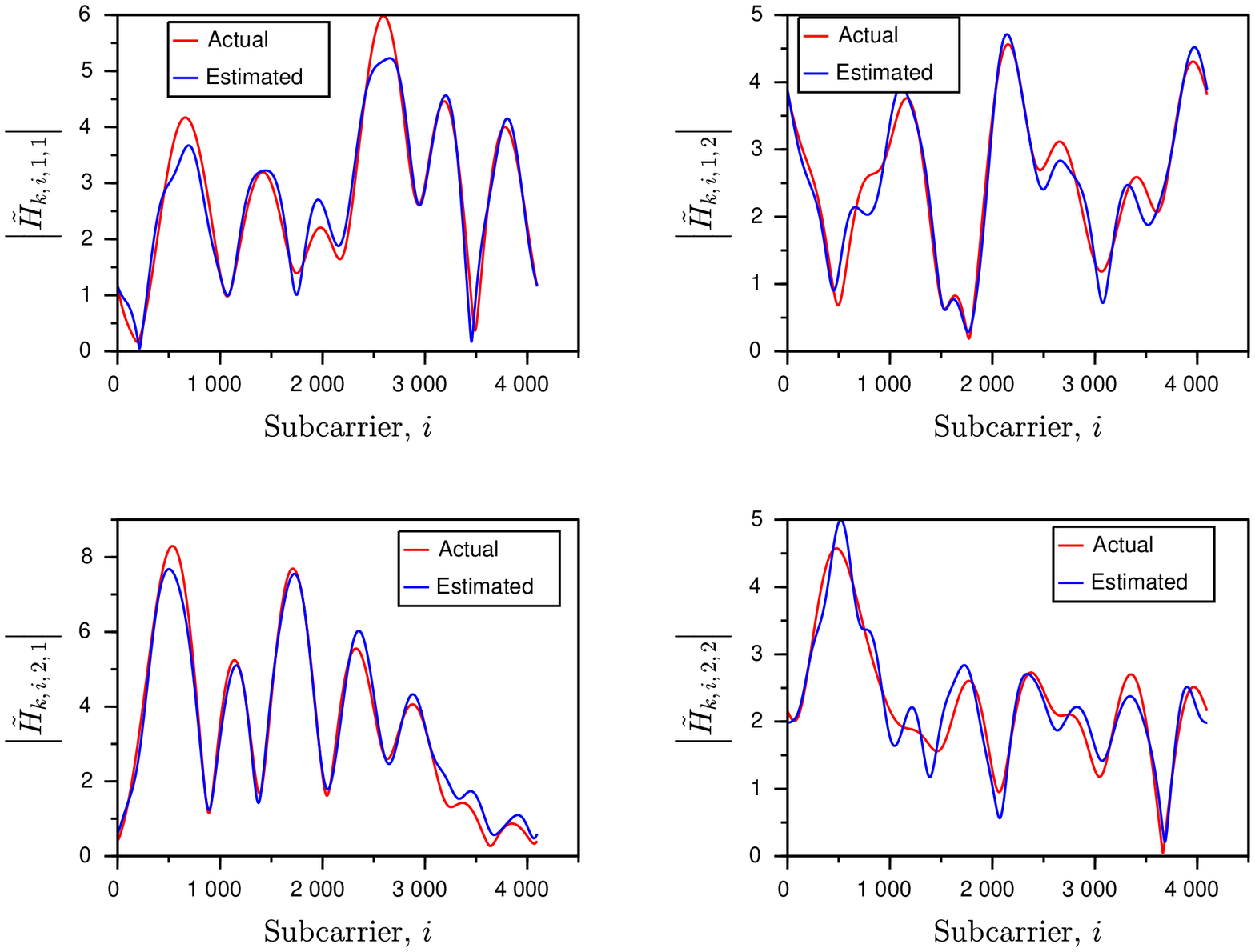}
\caption{Magnitude spectrum of estimated and actual channel, $L_p=4096$.}
\label{Fig:DMT36_E_Chest_0DB_Lp4096}
\end{figure*}
The magnitude spectrum of the actual and estimated channel for various
preamble lengths are shown in Figs.~\ref{Fig:DMT36_E_Chest_0DB_Lp512},
\ref{Fig:DMT36_E_Chest_0DB_Lp1024} and \ref{Fig:DMT36_E_Chest_0DB_Lp4096} for
$N_t=N_r=2$ and 0 dB average SNR per bit. Note that
$\tilde H_{k,\, i,\, n_r,\, n_t}$ denotes the
$L_d$-point discrete Fourier transform (DFT) of
$\tilde h_{k,\, n,\, n_r,\, n_t}$ in (\ref{Eq:Pap9_Eq3}).
\subsection{Fine Frequency Offset Estimation}
\label{SSec:Fine_FOFF_Est}
The fine frequency offset estimate is obtained using the following rule:
choose that value of time instant $m$ and frequency offset $\nu_{k,\, f}$
which maximizes:
\begin{eqnarray}
\label{Eq:Pap9_Eq25}
\left|
\left(
\tilde r_{k,\, m,\, n_r}
\,
\mathrm{e}^{-\mathrm{j}\,(\hat\omega_k+\nu_{k,\, f}) m}
\right)
\star
\tilde y_{1,\, k,\, L_2-1-m,\, n_r,\, n_t}^*
\right|
\end{eqnarray}
where
\begin{eqnarray}
\label{Eq:Pap9_Eq26}
 L_2                                 & = &  L_{hr} + L_p - 1   \nonumber  \\
\hat y_{1,\, k,\, m,\, n_r,\, n_t}   & = & \tilde s_{1,\, m,\, n_t}
                                           \star
                                           \hat h_{k,\, m,\, n_r,\, n_t}
\end{eqnarray}
where $\hat h_{k,\, m,\, n_r,\, n_t}$ is obtained from (\ref{Eq:Pap9_Eq21}).
The fine frequency offset
estimate ($\hat\nu_{k,\, f}(n_r,\, n_t)$) is obtained by dividing the interval
$[\hat\omega_k-0.005,\, \hat\omega_k+0.005]$ radian ($\hat\omega_k$ is given
in (\ref{Eq:Pap9_Eq19})) into $B_2=64$ frequency bins
\cite{Vasu_SIVP10}. The reason for choosing 0.005 radian can be traced to
Figure~5 of \cite{DBLP:journals/corr/Vasudevan15a}.
We find that the maximum error in the coarse
estimate of the frequency offset is approximately 0.004 radian over
$10^4$ frames. Thus the probability that the maximum error exceeds 0.005
radian is less than $10^{-4}$. However, from Table~\ref{Tbl:FOFF_RMS} in this
paper, we note that the maximum error in the frequency offset is
$2.4\times 10^{-2}$ radians for $L_p=512$, and $1.1\times 10^{-2}$ for
$L_p=1024$, both of which are larger than 0.005 radian. By observing this
trend, we expect that for larger values of $L_p$, say $L_p=4096$, the maximum
error in the coarse frequency offset estimate would be less than 0.005 radians.
Increasing $L_p$ would also imply an increase in $L_d$, for the same
throughput (see (\ref{Eq:Pap9_Throughput_Eq1})).
The average value of the fine frequency offset estimate is given by:
\begin{eqnarray}
\label{Eq:Pap9_Eq27}
\hat\omega_{k,\, f} = \frac{\sum_{n_r=1}^{N_r}\sum_{n_t=1}^{N_t}
                      \hat\nu_{k,\, f}(n_r,\, n_t)}
                      {N_r N_t}.
\end{eqnarray}
\subsection{Super Fine Frequency Offset Estimation}
\label{SSec:Super_Fine_FOFF_Est}
The fine frequency offset estimate in (\ref{Eq:Pap9_Eq27}) is still
inadequate for turbo decoding and data detection when $L_d \gg L_p$
\cite{6663392}. Note that the residual frequency offset is equal to:
\begin{eqnarray}
\label{Eq:Pap9_Eq28}
\omega_k - \hat\omega_k-\hat\omega_{k,\, f}.
\end{eqnarray}
This residual frequency offset is estimated by interpolating the FFT
output and performing postamble matched filtering at the receiver
\cite{Vasudevan2015,DBLP:journals/corr/Vasudevan15a}. If the
interpolation factor is $I$, then the FFT size is $IL_d$ (interpolation
in the frequency domain is achieved by zero-padding the FFT input in the time
domain, and then taking the $IL_d$-point FFT). Let
\begin{eqnarray}
\label{Eq:Pap9_Eq29}
m_{2,\, k} = m_{1,\, k} + L_p + L_{cs}
\end{eqnarray}
where $m_{1,\, k}$ is defined in (\ref{Eq:Pap9_Eq18_0}). Once again, we
drop the subscript $k$ from $m_{2,\, k}$ and refer to it as $m_2$.
Define the FFT input in the time domain as:
\begin{eqnarray}
\label{Eq:Pap9_Eq30}
\tilde\mathbf{r}_{k,\, m_2,\, n_r} =
\left[
\begin{array}{ccc}
\tilde r_{k,\, m_2,\, n_r} & \ldots & \tilde r_{k,\, m_2+L_d-1,\, n_r}
\end{array}
\right]^T
\end{eqnarray}
which is the data part of the received signal in (\ref{Eq:Pap9_Eq5}) for the
$k^{th}$ frame and receive antenna $n_r$, assumed to have the residual
frequency offset given by (\ref{Eq:Pap9_Eq28}). The output of the $IL_d$-point
FFT of $\tilde\mathbf{r}_{k,\, m_2,\, n_r}$ in (\ref{Eq:Pap9_Eq30}) is denoted
by
\begin{eqnarray}
\label{Eq:Pap9_Eq31}
\tilde R_{k,\, i,\, n_r} =
\sum_{n=0}^{L_d - 1}
\tilde r_{k,\, m_2+n,\, n_r}
\mathrm{e}^{-\mathrm{j}\, 2\pi i n/(IL_d)}
\end{eqnarray}
for $0 \le i \le I L_d - 1$.

The coefficients of the postamble matched filter is obtained as follows
\cite{Vasudevan2015,DBLP:journals/corr/Vasudevan15a}. Define
\begin{eqnarray}
\label{Eq:Pap9_Eq32}
\tilde G_{k,\, i,\, n_r}'' = \sum_{n_t=1}^{N_t}
                             \hat H_{k,\, i_3,\, n_r,\, n_t}
                              S_{k,\, 3,\, i,\, n_t}
                             \quad
                             \mbox{for $i_0 \le i \le i_1$}
\end{eqnarray}
where $\hat H_{k,\, i,\, n_r,\, n_t}$ is the $L_d$-point FFT of the
channel estimate in (\ref{Eq:Pap9_Eq21}), and
\begin{eqnarray}
\label{Eq:Pap9_Eq33}
i_0 & = & B + L_{d2}                        \nonumber  \\
i_1 & = & i_0 + L_o - 1                     \nonumber  \\
i_3 & = & B + \pi(i-B)
\end{eqnarray}
where $\pi(\cdot)$ is the data interleaver
map, $B$, $L_{d2}$ and $L_o$ are the lengths of the buffer, data, and
postamble, respectively, as shown in Figure~\ref{Fig:Enh_Frame1}(b). Let
\begin{eqnarray}
\label{Eq:Pap9_Eq34}
\tilde G_{k,\, i_3,\, n_r}' =
\left
\{
\begin{array}{ll}
\tilde G_{k,\, i,\, n_r}'' & \mbox{for $i_0 \le i \le i_1$}\\
 0                         & \mbox{otherwise}
\end{array}
\right.
\end{eqnarray}
where $0 \le i_3 \le L_d -1$, the relation between $i_3$ and $i$ is given
in (\ref{Eq:Pap9_Eq33}). Next, we perform interpolation:
\begin{eqnarray}
\label{Eq:Pap9_Eq35}
\tilde G_{k,\, i_4,\, n_r} =
\left
\{
\begin{array}{ll}
\tilde G_{k,\, i,\, n_r}' & \mbox{for $0 \le i \le L_d-1$}\\
 0                        & \mbox{otherwise}
\end{array}
\right.
\end{eqnarray}
where $0 \le i_4 \le IL_d-1$ and $i_4=iI$. Finally, the postamble matched
filter is $\tilde G_{k,\, IL_d-1-i,\, n_r}^*$, which is convolved with
$\tilde R_{k,\, i,\, n_r}$ in (\ref{Eq:Pap9_Eq31}).
Note that due to the presence of the cyclic prefix, any residual
frequency offset in the time domain, manifests as a circular
shift in the frequency domain. The purpose of the postamble matched filter
is to capture this shift. The role of the buffer symbols is explained in
\cite{Vasudevan2015,DBLP:journals/corr/Vasudevan15a}. Assume that the peak of
the postamble matched filter output occurs at $m_{3,\,k}(n_r)$. Ideally, in
the absence of noise and frequency offset
\begin{eqnarray}
\label{Eq:Pap9_Eq35_1}
m_{3,\, k}(n_r)=IL_d-1.
\end{eqnarray}
In the presence of the frequency
offset, the peak occurs to the left or right of $IL_d-1$. The average superfine
estimate of the residual frequency offset is given by:
\begin{eqnarray}
\label{Eq:Pap9_Eq36}
\hat \omega_{k,\, sf} =  2\pi/(I L_d N_r)
                        \sum_{n_r=1}^{N_r}
                        \left[
                         m_{3,\, k}(n_r) - IL_d + 1
                        \right].
\end{eqnarray}
\subsection{Noise Variance Estimation}
\label{SSec:Noise_Var_Est}
The noise variance is estimated as follows, for the purpose of turbo decoding:
\begin{eqnarray}
\label{Eq:Pap9_Eq37}
\hat
\sigma^2_w =
\frac{1}{2 L_p N_r}
\sum_{n_r=1}^{N_r}
\left(
\tilde\mathbf{r}_{k,\, m_1,\, n_r} -
\sum_{n_t=1}^{N_t}
\tilde\mathbf{s}_{5,\, n_t}
\hat\mathbf{h}_{k,\, n_r,\, n_t}
\right)^H      \nonumber  \\
\left(
\tilde\mathbf{r}_{k,\, m_1,\, n_r} -
\sum_{n_t=1}^{N_t}
\tilde\mathbf{s}_{5,\, n_t}
\hat\mathbf{h}_{k,\, n_r,\, n_t}
\right).
\end{eqnarray}
\subsection{Turbo Decoding}
\label{SSec:Turbo_Dec}
In this section, we assume that the frequency offset has been perfectly
canceled, that is, $\tilde\mathbf{r}_{k,\, m_2,\, n_r}$
in (\ref{Eq:Pap9_Eq30}) contains no frequency offset. The output of the
$L_d$-point FFT of $\tilde\mathbf{r}_{k,\, m_2,\, n_r}$ for the $k^{th}$ frame
is given by:
\begin{eqnarray}
\label{Eq:Pap9_Eq37_1}
\tilde R_{k,\, i,\, n_r} = \sum_{n_t=1}^{N_t}
                           \tilde H_{k,\, i,\, n_r,\, n_t}
                            S_{k,\, 3,\, i,\, n_t} +
                           \tilde W_{k,\, i,\, n_r}
\end{eqnarray}
for $0 \le i \le L_d - 1$, where $\tilde H_{k,\, i,\, n_r,\, n_t}$ is the
$L_d$-point FFT of $\tilde h_{k,\, n,\, n_r,\, n_t}$ and
$\tilde W_{k,\, i,\, n_r}$ is
the $L_d$-point FFT of $\tilde w_{k,\, n,\, n_r}$. It can be shown that
\cite{6663392,DBLP:journals/corr/Vasudevan15a}
\begin{eqnarray}
\label{Eq:Pap9_Eq37_2}
\frac{1}{2}
 E
\left[
\left|
\tilde W_{k,\, i,\, n_r}
\right|^2
\right] & = & L_d \sigma^2_w      \nonumber  \\
\frac{1}{2}
 E
\left[
\left|
\tilde H_{k,\, i,\, n_r,\, n_t}
\right|^2
\right] & = & L_h \sigma^2_f.
\end{eqnarray}
The received signal in (\ref{Eq:Pap9_Eq37_1}), for the $k^{th}$ frame and
$i^{th}$ subcarrier, can be written in matrix form as follows:
\begin{eqnarray}
\label{Eq:Pap9_Eq37_3}
\tilde \mathbf{R}_{k,\, i} =
			   \tilde \mathbf{H}_{k,\, i}
			   \mathbf{S}_{k,\, 3,\, i} +
			   \tilde \mathbf{W}_{k,\, i}
\end{eqnarray}
where $\tilde\mathbf{R}_{k,\, i}$ is the $N_r\times 1$ received signal vector,
$\tilde\mathbf{H}_{k,\, i}$ is the $N_r\times N_t$ channel matrix,
$\mathbf{S}_{k,\, 3,\, i}$ is the $N_t\times 1$ symbol vector and
$\tilde\mathbf{W}_{k,\, i}$ is the $N_r\times 1$ noise vector.

The generating matrix of each of the constituent encoders is given by (41)
in \cite{DBLP:journals/corr/Vasudevan15a}, and is repeated here for
convenience:
\begin{eqnarray}
\label{Eq:Pap9_Eq38}
\mathbf{G}(D) =
\left[
\begin{array}{cc}
 1 & \frac{\displaystyle 1+D^2}{\displaystyle 1+D+D^2}
\end{array}
\right].
\end{eqnarray}
For the purpose of turbo decoding, we consider the case where $N_r=N_t=2$.
The details of turbo decoding can be found in
\cite{DBLP:journals/corr/Vasudevan15a}, and will not be discussed here.
Suffices to say that corresponding to the transition from state $m$ to state
$n$, at decoder 1, for the $k^{th}$ frame, at time $i$, we define
(for $0 \le i \le L_{d2} - 1$):
\begin{eqnarray}
\label{Eq:Pap9_Turbo_Eq10}
\gamma_{1,\, k,\, i,\, m,\, n}
                          = \exp
                            \left(
                            -Z_{1,\, k,\, i,\, m,\, n}/
                            \left(
                             2L_d\hat\sigma^2_w
                            \right)
                            \right)
\end{eqnarray}
where $Z_{1,\, k,\, i,\, m,\, n}$ is given by
\begin{eqnarray}
\label{Eq:Pap9_Turbo_Eq10_0}
\min_{\mathrm{all}\,\,  S_{m,\, n,\, 2}}
\sum_{n_r=1}^{2}
\left|
\tilde R_{k,\, i,\, n_r} -
\sum_{n_t=1}^{2}
\hat H_{k,\, i,\, n_r,\, n_t}
 S_{m,\, n,\, n_t}
\right|^2
\end{eqnarray}
where $S_{m,\, n,\, n_t}$ denotes the QPSK symbol corresponding to the
transition from state $m$ to state $n$ in the trellis, at transmit antenna
$n_t$. Observe that $\hat\sigma^2_w$ is the estimate of $\sigma^2_w$ obtained
from (\ref{Eq:Pap9_Eq37}). Observe that the minimization in
(\ref{Eq:Pap9_Turbo_Eq10_0}) is over all possible QPSK symbols, at $n_t=2$
and index $i$. Similarly, for the transition from state $m$ to
state $n$, at decoder 2, for the $k^{th}$ frame, at time $i$, we
define (for $0 \le i \le L_{d2}-1$):
\begin{eqnarray}
\label{Eq:Pap9_Turbo_Eq10_1}
\gamma_{2,\, k,\, i,\, m,\, n}
                          = \exp
                            \left(
                            -Z_{2,\, k,\, i,\, m,\, n}/
                            \left(
                             2L_d\hat\sigma^2_w
                            \right)
                            \right)
\end{eqnarray}
where $Z_{2,\, k,\, i,\, m,\, n}$ is given by
\begin{eqnarray}
\label{Eq:Pap9_Turbo_Eq10_2}
\min_{\mathrm{all}\,\,  S_{m,\, n,\, 1}}
\sum_{n_r=1}^{2}
\left|
\tilde R_{k,\, i,\, n_r} -
\sum_{n_t=1}^{2}
\hat H_{k,\, i,\, n_r,\, n_t}
 S_{m,\, n,\, n_t}
\right|^2.
\end{eqnarray}
Now, (\ref{Eq:Pap9_Turbo_Eq10}) and (\ref{Eq:Pap9_Turbo_Eq10_1}) are used in
the forward and backward recursions of the BCJR algorithm
\cite{DBLP:journals/corr/Vasudevan15a}.
\subsection{Summary of the Receiver Algorithms}
\label{SSec:Rx_Algo_Sum}
The receiver algorithms are summarized as follows:
\begin{enumerate}
 \item Estimate the start-of-frame and the frequency offset (coarse)
       using (\ref{Eq:Pap9_Eq16}), for each receive antenna. Obtain the
       average value of the frequency offset ($\hat\omega_k$) using
       (\ref{Eq:Pap9_Eq19}).
 \item Cancel the frequency offset by multiplying
       $\tilde r_{k,\, n,\, n_r}$ in (\ref{Eq:Pap9_Eq5}) by
       $\mathrm{e}^{-\mathrm{j}\,\hat\omega_k n}$, and estimate the channel
       using (\ref{Eq:Pap9_Eq21}), for each $n_r$ and $n_t$.
 \item Obtain $\tilde y_{1,\, k,\, m,\, n_r,\, n_t}$ from (\ref{Eq:Pap9_Eq26})
       and the fine frequency offset using (\ref{Eq:Pap9_Eq27}).
 \item Cancel the frequency offset by multiplying
       $\tilde r_{k,\, n,\, n_r}$ in (\ref{Eq:Pap9_Eq5}) by
       $\mathrm{e}^{-\mathrm{j}(\hat\omega_k+\hat\omega_{k,\, f}) n}$, and
       estimate the channel again using (\ref{Eq:Pap9_Eq21}), for each $n_r$
       and $n_t$.
 \item Obtain the average superfine frequency offset estimate using
       (\ref{Eq:Pap9_Eq36}). Cancel the offset by multiplying
       $\tilde r_{k,\, n,\, n_r}$ in (\ref{Eq:Pap9_Eq5}) by
       $\mathrm{e}^{-\mathrm{j}(\hat\omega_k+\hat\omega_{k,\, f}+
       \hat\omega_{k,\, sf}) n}$.
 \item Obtain the noise variance estimate from (\ref{Eq:Pap9_Eq37}).
 \item Take the $L_d$-point FFT of $\tilde\mathbf{r}_{k,\, m2,\, n_r}$
       and perform turbo decoding.
\end{enumerate}

\subsection{Simulation Results}
\label{SSec:Results}
In this section, we present the simulation results for the proposed turbo
coded MIMO OFDM system with $N_t=N_r=2$. The SNR per bit is defined in
(\ref{Eq:CCT_Eq12}). Note that one data bit (two coded QPSK symbols) is sent
simultaneously from two transmit antennas. Hence, the number of data bits
sent from each transmit antenna is $\kappa=0.5$, as given in
(\ref{Eq:CCT_Eq12}). We have also assumed that $\sigma^2_f=0.5$. The frame
parameters are summarized in Table~\ref{Tbl:Sim_Param}. The maximum number of
frames simulated is $2.2\times 10^4$, at an average SNR per bit of 6.5 dB.
Each simulation run is over $10^3$ frames. Hence, the maximum number of
independently seeded simulation runs is 22.
\begin{table}[tbh]
\centering
\caption{Frame parameters.}
\input{sim_param.pstex_t}
\label{Tbl:Sim_Param}
\end{table}

The throughput is defined as
\cite{Vasudevan2015,DBLP:journals/corr/Vasudevan15a}:
\begin{eqnarray}
\label{Eq:Pap9_Throughput_Eq1}
\mathscr{T} = \frac{L_{d2}}{L_d+L_p+L_{cp}+L_{cs}}.
\end{eqnarray}
\begin{table}[tbh]
\centering
\caption{Throughput.}
\input{throughput.pstex_t}
\label{Tbl:Throughput}
\end{table}
The throughput of various frame configurations is given in
Table~\ref{Tbl:Throughput}.
\begin{figure}[tbh]
\centering
\input{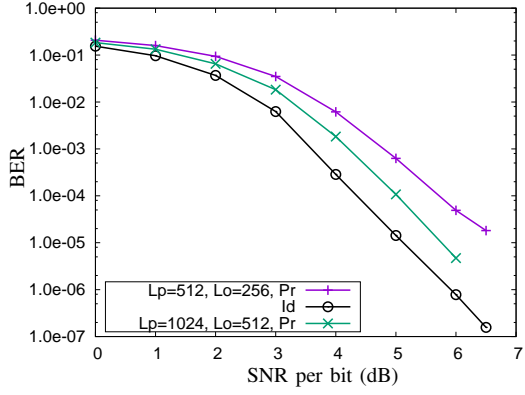}
\caption{BER simulation results.}
\label{Fig:MIMO_2_2_BER}
\end{figure}
The BER simulation results for the turbo coded MIMO OFDM system with
$N_t=N_r=2$ is shown in Figure~\ref{Fig:MIMO_2_2_BER}.
Here ``Id'' denotes the ideal receiver. For the practical
receivers (``Pr''), the interpolation factor for superfine
frequency offset estimation is $I=16$. The
practical receiver with $L_p=1024,\, L_o=512$ attains a BER of $10^{-5}$
at an SNR per bit of 5.5 dB, which is 1 dB better than the receiver
with $L_p=512,\, L_o=256$. This is due to the fact that the variance of the
channel estimation error with $L_p=512$ is twice that of $L_p=1024$ (see
(\ref{Eq:Pap9_Eq24})). This difference in the variance of the channel
estimation error affects the turbo decoding process. Moreover, the
practical receiver in Figure~\ref{Fig:MIMO_2_2_BER} with $L_p=1024,\, L_o=512$
is 2.5 dB better than the practical receiver with one transmit and two receive
antennas in Figure~10 of \cite{Vasudevan2015}.
\begin{table}[tbh]
\centering
\caption{Probability of frame erasure.}
\input{feer_0db.pstex_t}
\label{Tbl:Feer_0dB}
\end{table}

The probability of frame erasure (this happens when (\ref{Eq:Pap9_Eq18}) is
not satisfied) at 0 dB SNR per bit is shown in Table~\ref{Tbl:Feer_0dB}.
Clearly, as $L_p$ increases, the probability of erasure decreases.
\begin{table*}[tbh]
\centering
\caption{Frequency offset estimation error.}
\input{foff_rms.pstex_t}
\label{Tbl:FOFF_RMS}
\end{table*}
Finally, the root mean square (RMS) and maximum frequency offset estimation
error in radians, at 0 dB SNR per bit, is given in Table~\ref{Tbl:FOFF_RMS}.

\section{Near Capacity Signaling}
\label{Sec:Near_Capacity}
In Sections~\ref{Sec:Sys_Model} and \ref{Sec:Receiver}, we had presented the
discrete-time algorithms for MIMO-OFDM. The inherent assumption in these
two sections was that all transmit antennas use the same carrier frequency. The
consequence of this assumption is that the signal at each receive antenna is a
linear combination of the symbols from all the transmit antennas, as given in
(\ref{Eq:Pap9_Eq37_1}) and (\ref{Eq:Pap9_Eq37_3}). This makes the estimation of
symbols, $\mathbf{S}_{k,\, 3,\, i}$ in (\ref{Eq:Pap9_Eq37_3}), complicated for
large values of $N_t$ and $N_r$ (massive MIMO). In this section, we assume that
distinct transmit antennas use different carrier frequencies. Thus, the signals
from distinct transmit antennas are orthogonal. To each transmit antenna, we
associate $N_r$ receive antennas, that are capable of receiving signals
from one particular transmit antenna. The total number of receive antennas
is now $N_r N_t$.

In order to restrict the transmitted signal spectrum,
it is desirable to have a lowpass filter (LPF) at the output of
the parallel-to-serial converter in Figure~\ref{Fig:Frame}, for each transmit
antenna. If we assume that the cut-off frequency of the LPF is $\pi/10$
radians and its transition bandwidth is $\pi/20$ radians, then the required
length of the linear-phase, finite impulse response (FIR) LPF with Hamming
window would be \cite{Proakis92}
\begin{eqnarray}
\label{Eq:Pap10_Near_Cap_Eq0}
8\pi/L_{\mathrm{LPF}} & = & \pi/20                         \nonumber  \\
\Rightarrow
L_{\mathrm{LPF}}      & = & 160.
\end{eqnarray}
Note that an infinite impulse response (IIR) filter could also be used.
However, it may have stability problems when the cut-off frequency of the LPF
is close to zero radians.
If the physical channel has 10 taps as given by (\ref{Eq:Pap9_Eq1_1}), then
the length of the equivalent channel as seen by the receiver would be:
\begin{eqnarray}
\label{Eq:Pap10_Near_Cap_Eq0_1}
L_h & = & L_{\mathrm{LPF}} + 10 - 1        \nonumber  \\
    & = & 160+10-1                         \nonumber  \\
    & = & 169.
\end{eqnarray}
The values of $L_p$ and $L_d$ in Figure~\ref{Fig:Enh_Frame1}(a) have to be
suitably increased to obtain a good
estimate of the channel (see (\ref{Eq:Pap9_Eq24})) and maintain a
high throughput (see \ref{Eq:Pap9_Throughput_Eq1}).
Let us denote the impulse response of the LPF by $p_n$. We assume that
$p_n$ is obtained by sampling the continuous-time impulse response $p(t)$
at a rate of $1/T_s$, where $T_s$ is defined in (\ref{Eq:Pap9_Eq1_1}). Note
that $p_n$ is real-valued \cite{Proakis92}. The discrete-time Fourier
transform (DTFT) of $p_n$ is \cite{Vasu_Book10,Vasu_Book16}:
\begin{eqnarray}
\label{Eq:Pap10_Near_Cap_Eq0_2}
\tilde P_{\mathscr{P}}(F) & = & \sum_{n=0}^{L_{\mathrm{LPF}}-1}
                                 p_n \mathrm{e}^{-\mathrm{j}\, 2\pi F nT_s}
                                \nonumber  \\
                          & = & \frac{1}{T_s}
                                \sum_{m=-\infty}^{\infty}
                                \tilde P(F-m/T_s)
\end{eqnarray}
where the subscript $\mathscr{P}$ denotes a periodic function, $F$ denotes the
frequency in Hz and $\tilde P(F)$ is the continuous-time
Fourier transform of $p(t)$. Observe that:
\begin{enumerate}
 \item the digital frequency $\omega$ in radians is given by
\begin{eqnarray}
\label{Eq:Pap10_Near_Cap_Eq0_3}
\omega = 2\pi F T_s
\end{eqnarray}
 \item $\tilde P_{\mathscr{P}}(F)$ is periodic with period $1/T_s$
 \item there is no aliasing in the second equation of
      (\ref{Eq:Pap10_Near_Cap_Eq0_2}), that is
\begin{eqnarray}
\label{Eq:Pap10_Near_Cap_Eq0_4}
\tilde P_{\mathscr{P}}(F) = \frac{\tilde P(F)}{T_s}
                            \qquad
                            \mbox{for $-\frac{1}{2T_s} < F < \frac{1}{2T_s}$}.
\end{eqnarray}
\end{enumerate}
Now, $\tilde s_{k,\, n,\, n_t}$ in Figure~\ref{Fig:Enh_Frame1}(a) is passed
through the LPF. Let us denote the LPF output by $\tilde v_{k,\, n,\, n_t}$.
After digital-to-analog (D/A) conversion, the continuous-time signal is
denoted by $\tilde v_{k,\, n_t}(t)$. The power spectral density of
$\tilde v_{k,\, n_t}(t)$ \cite{Vasu_Book10,Vasu_Book16}
\begin{eqnarray}
\label{Eq:Pap10_Near_Cap_Eq0_5}
S_{\tilde v}(F) = \frac{1}{T_s} \cdot
                  \frac{\sigma^2_s}{2} \cdot
                  \left|
                  \tilde P(F)
                  \right|^2
\end{eqnarray}
where we have assumed that the samples of $\tilde s_{k,\, n,\, n_t}$
are uncorrelated with variance $\sigma^2_s$ given in (\ref{Eq:Pap9_Eq4_2}).
Thus the one-sided bandwidth of the complex baseband signal
$\tilde v_{k,\, n_t}(t)$ is $1/(20T_s)$ Hz, for an LPF with cut-off frequency
$\pi/10$ radians, since $1/T_s$ corresponds to $2\pi$ radians. Thus, the
passband signal spectrum from a single transmit
antenna would have a two-sided bandwidth of $1/(10T_s)$ Hz.

The frame structure is given by
Figure~\ref{Fig:Enh_Frame1}. The average power of the preamble in the time
domain must be equal to that of the data, as given by (\ref{Eq:Pap9_Eq4_2}).
Due to the use of different carrier frequencies for distinct transmit
antennas, the same preamble pattern can be used for all the transmit
antennas. Therefore, the subscript $n_t$ can be dropped from the preamble
signal, both in the time and frequency domain, in Figure~\ref{Fig:Enh_Frame1}(a)
and (\ref{Eq:Pap9_Eq4_1}). There are also no zero-valued
preamble symbols in the frequency domain, that is \cite{6663392}
\begin{eqnarray}
\label{Eq:Pap10_Preamb_Eq1}
S_{1,\, i}\in \sqrt{L_p/L_d}\,(\pm 1 \pm \mathrm{j})
\end{eqnarray}
for $0\le i \le L_p-1$.
\begin{figure}[tbh]
\centering
\input{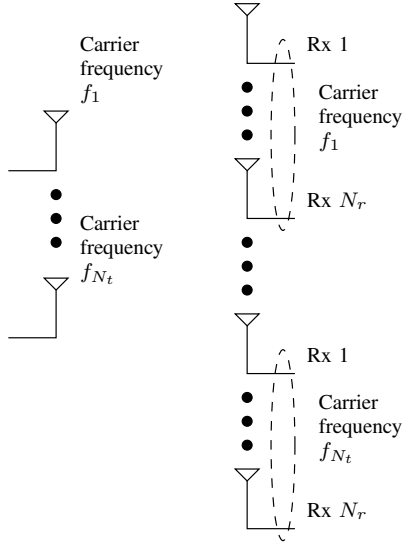}
\caption{Near capacity signaling.}
\label{Fig:Near_Capacity}
\end{figure}
The block diagram of the system for near capacity signaling is shown in
Figure~\ref{Fig:Near_Capacity}. The received signal vector at the output of
the FFT for the $N_r$ antennas associated with the transmit antenna $n_t$,
for the $k^{th}$ frame and $i^{th}$ ($0\le i \le L_d-1$) subcarrier is:
\begin{eqnarray}
\label{Eq:Pap10_Near_Cap_Eq1}
\tilde \mathbf{R}_{k,\, i,\, n_t} =
			   \tilde \mathbf{H}_{k,\, i,\, n_t}
			    S_{k,\, 3,\, i,\, n_t} +
			   \tilde \mathbf{W}_{k,\, i,\, n_t}
\end{eqnarray}
where $S_{k,\,3,\, i,\, n_t}$ is given in (\ref{Eq:Pap9_Eq4_1}),
$\tilde\mathbf{R}_{k,\, i,\, n_t}$, $\tilde\mathbf{H}_{k,\, i,\, n_t}$
and $\tilde\mathbf{W}_{k,\, i,\, n_t}$ are $N_r\times 1$ vectors given by:
\begin{eqnarray}
\label{Eq:Pap10_Near_Cap_Eq2}
\tilde \mathbf{R}_{k,\, i,\, n_t}
& = &
\left[
\begin{array}{ccc}
\tilde R_{k,\, i,\, n_t,\, 1} & \ldots &
\tilde R_{k,\, i,\, n_t,\, N_r}
\end{array}
\right]^T                                          \nonumber  \\
\tilde \mathbf{H}_{k,\, i,\, n_t}
& = &
\left[
\begin{array}{ccc}
\tilde H_{k,\, i,\, n_t,\, 1} & \ldots &
\tilde H_{k,\, i,\, n_t,\, N_r}
\end{array}
\right]^T                                          \nonumber  \\
\tilde \mathbf{W}_{k,\, i,\, n_t}
& = &
\left[
\begin{array}{ccc}
\tilde W_{k,\, i,\, n_t,\, 1} & \ldots &
\tilde W_{k,\, i,\, n_t,\, N_r}
\end{array}
\right]^T.                                         \nonumber  \\
\end{eqnarray}
Similar to (\ref{Eq:Pap9_Eq37_2}), it can be shown that for
$1\le l\le N_r$
\begin{eqnarray}
\label{Eq:Pap10_Near_Cap_Eq3}
\frac{1}{2}
 E
\left[
\left|
\tilde W_{k,\, i,\, n_t,\, l}
\right|^2
\right] & = & L_d \sigma^2_w      \nonumber  \\
\frac{1}{2}
 E
\left[
\left|
\tilde H_{k,\, i,\, n_t,\, l}
\right|^2
\right] & = & L_h \sigma^2_f.
\end{eqnarray}
The synchronization and channel estimation algorithms are identical to
that given in Section~\ref{Sec:Receiver} with $N_t=1$.

In the turbo decoding operation we assume that $N_t=2$. The generating
matrix of the constituent encoders is given by (\ref{Eq:Pap9_Eq38}). For
decoder 1 and $0 \le i \le L_{d2} -1$, we define \cite{Vasudevan2015}:
\begin{eqnarray}
\label{Eq:Turbo_Eq38_4}
\gamma_{1,\, k,\, i,\, m,\, n} =
\prod_{l=1}^{N_r}
\gamma_{1,\, k,\, i,\, m,\, n,\, l}
\end{eqnarray}
where
\begin{eqnarray}
\label{Eq:Turbo_Eq38_5}
\gamma_{1,\, k,\, i,\, m,\, n,\, l} =
                            \exp
                            \left[-
                            \frac{
                            \left|
                            \tilde R_{k,\, i,\, 1,\, l}-
                            \hat H_{k,\, i,\, 1,\, l}
                             S_{m,\, n}
                            \right|^2}{2L_d\hat\sigma_w^2}
                            \right]
\end{eqnarray}
where $\hat\sigma_w^2$ is the average estimate of the noise variance over
all the $N_r$ diversity arms, as given by (\ref{Eq:Pap9_Eq37}) with $N_t=1$,
and $S_{m,\, n}$ is the QPSK symbol
corresponding to the transition from state $m$ to $n$ in the encoder trellis.
Similarly at decoder 2, for $0 \le i \le L_{d2}-1$, we have:
\begin{eqnarray}
\label{Eq:Turbo_Eq38_6}
\gamma_{2,\, k,\, i,\, m,\, n} =
\prod_{l=1}^{N_r}
\gamma_{2,\, k,\, i,\, m,\, n,\, l}
\end{eqnarray}
where
\begin{eqnarray}
\label{Eq:Turbo_Eq38_7}
\gamma_{2,\, k,\, i,\, m,\, n,\, l} =
                            \exp
                            \left[-
                            \frac{
                            \left|
                            \tilde R_{k,\, i,\, 2,\, l}-
                            \hat H_{k,\, i,\, 2,\, l}
                             S_{m,\, n}
                            \right|^2}{2L_d\hat\sigma_w^2}
                            \right].
\end{eqnarray}
\begin{figure}[tbh]
\centering
\includegraphics[scale=0.6]{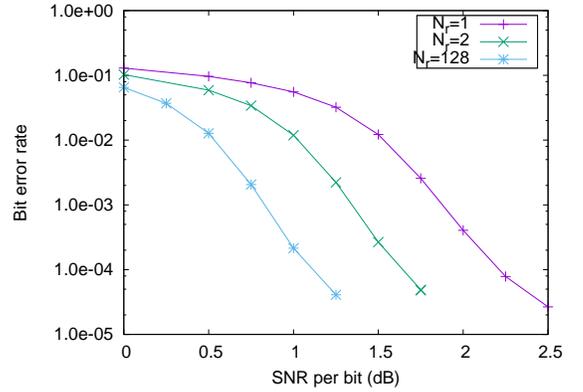}
\caption{BER results for near capacity signaling with $N_t=2$.}
\label{Fig:DMT38_BER}
\end{figure}

The simulation results assuming an ideal coherent receiver is given in
Figure~\ref{Fig:DMT38_BER}, for $L_d=4096$, $L_p=L_o=B=L_{cp}=L_{cs}=0$
(the preamble, postamble, buffer, cyclic prefix or suffix is not required,
since this is an ideal coherent receiver), $N_t=2$ and different
values of $N_r$.
\begin{table}[tbh]
\centering
\caption{Number of simulation runs for various $N_r$.}
\input{sim_run.pstex_t}
\label{Tbl:Sim_Run}
\end{table}

The maximum number of independently seeded simulation runs for various $N_r$
and SNR per bit is given in Table~\ref{Tbl:Sim_Run}. For lower values of the
SNR per bit, the number of runs is less than the maximum. Each run is over
$10^3$ frames. The time taken for one run using Scilab on an i5 processor
is also given in Table~\ref{Tbl:Sim_Run}. The total time taken to obtain
Figure~\ref{Fig:DMT38_BER} is
approximately three months. The channel coefficients
$\tilde H_{k,\, i,\, n_t,\, l}$ in (\ref{Eq:Pap10_Near_Cap_Eq2})
are assumed to be complex Gaussian and independent over $i$ and $l$, that is
\begin{eqnarray}
\label{Eq:Pap10_Near_Cap_Eq3_1}
\frac{1}{2}
 E
\left[
\tilde H_{k,\, i_1,\, n_t,\, l}
\tilde H_{k,\, i_2,\, n_t,\, l}^*
\right] & = & L_h \sigma^2_f\delta_K(i_1-i_2)           \nonumber  \\
\frac{1}{2}
 E
\left[
\tilde H_{k,\, i,\, n_t,\, l_1}
\tilde H_{k,\, i,\, n_t,\, l_2}^*
\right] & = & L_h \sigma^2_f\delta_K(l_1-l_2).          \nonumber  \\
\end{eqnarray}
The average SNR per bit is given by the second
equation in (\ref{Eq:CCT_Eq12}) with
\begin{eqnarray}
\label{Eq:Pap10_Near_Cap_Eq4}
P_{\mathrm{av}} & = & 2         \nonumber  \\
L_h \sigma^2_f  & = & 0.5       \nonumber  \\
        \kappa  & = & 0.5.
\end{eqnarray}
The turbo decoder uses eight iterations.
Observe that in Figure~\ref{Fig:DMT38_BER}, we obtain a BER of $2\times 10^{-5}$
at an average SNR per bit of just 2.5 dB, for $N_r=1$.
It is also clear from Figure~\ref{Fig:DMT38_BER} that increasing $N_r$ follows
the law of diminishing returns. In fact there is only 1.25 dB difference
in the average SNR per bit, between $N_r=1$ and $N_r=128$, at a BER
of $2\times 10^{-5}$. The slight change in slope at a BER of $2\times 10^{-5}$
is probably because the BER needs to averaged over more number of simulation
runs. We do not expect an ideal coherent detector to exhibit an error floor.

It is interesting to compare the average SNR per bit definitions given by
(\ref{Eq:Pap10_Near_Cap_Eq4}) and (\ref{Eq:CCT_Eq12}) for $N_r=1$ in this
paper, with (38) in \cite{6663392}. Observe that both definitions are
identical. However, in Figure~\ref{Fig:DMT38_BER}, we obtain
a BER of $2\times 10^{-5}$ at an average SNR per bit of 2.5 dB, whereas in
Figure~7 of \cite{6663392} we obtain a similar BER at 8 dB average SNR per bit,
for the ideal receiver. What could be the reason for this difference?
Simply stated, in this section we have assumed a $4096$-tap channel
(see the first equation in
(\ref{Eq:Pap10_Near_Cap_Eq3_1}) and equation (37) in \cite{6663392} with
$L_h=L_d$). However, in \cite{6663392} we have considered a 10-tap channel.
This is further explained below.

\begin{enumerate}
 \item In this section, the SNR per bit for the $k^{th}$ frame and $N_r=1$ is
       proportional to (see also (22) in \cite{Vasudevan2015})
\begin{eqnarray}
\label{Eq:Pap10_Near_Cap_Eq5}
\mbox{SNR}_{k,\,\mathrm{bit},\, 1}
                              \propto
                              \frac{1}{L_d}
                              \sum_{i=0}^{L_d-1}
                              \left|
                              \tilde H_{k,\, i,\, n_t,\, 1}
                              \right|^2
\end{eqnarray}
       where the subscript $1$ in $\mbox{SNR}_{k,\,\mathrm{bit},\, 1}$
       denotes case (1) and $\tilde H_{k,\, i,\, n_t,\, 1}$ is defined in
       (\ref{Eq:Pap10_Near_Cap_Eq2}). Note that $\tilde H_{k,\, i,\, n_t,\, 1}$
       is a zero-mean Gaussian random variable which is independent over $i$
       and variance given by (\ref{Eq:Pap10_Near_Cap_Eq3}). Moreover, the
       right hand side of (\ref{Eq:Pap10_Near_Cap_Eq5}) gives the estimate of
       the variance of $\tilde H_{k,\, i,\, n_t,\, 1}$. Let us now compute the
       variance of the estimate of the variance in
       (\ref{Eq:Pap10_Near_Cap_Eq5}), that is
\begin{eqnarray}
\label{Eq:Pap10_Near_Cap_Eq6}
\sigma_1^2 =
 E
\left[
\left(
\frac{1}{L_d}
\sum_{i=0}^{L_d-1}
\left|
\tilde H_{k,\, i,\, n_t,\, 1}
\right|^2 - 2L_h\sigma^2_f
\right)^2
\right]
\end{eqnarray}
       where we have used (\ref{Eq:Pap10_Near_Cap_Eq3}). It can be shown that
\begin{eqnarray}
\label{Eq:Pap10_Near_Cap_Eq7}
\sigma_1^2 = \frac{\sigma_H^4}{L_d} = \frac{4L_h^2\sigma_f^4}{L_d}
\end{eqnarray}
       where for $0\le i \le L_d-1$
\begin{eqnarray}
\label{Eq:Pap10_Near_Cap_Eq8}
 E
\left[
\left|
\tilde H_{k,\, i,\, n_t,\, 1}
\right|^2
\right]    & = &  2L_h \sigma_f^2                  \nonumber \\
           & \stackrel{\Delta}{=}
               & \sigma_H^2                        \nonumber  \\
H_{k,\, i,\, n_t,\, 1,\, I} +
\mathrm{j}\,
H_{k,\, i,\, n_t,\, 1,\, Q}
           & = & \tilde H_{k,\, i,\, n_t,\, 1}
                                                   \nonumber  \\
 E
\left[
 H_{k,\, i,\, n_t,\, 1,\, I}^2
\right]    & = & \sigma_H^2/2                      \nonumber  \\
           & \stackrel{\Delta}{=}
               & \sigma_{H,\, 1}^2                 \nonumber  \\
 E
\left[
 H_{k,\, i,\, n_t,\, 1,\, Q}^2
\right]    & = & \sigma_H^2/2                      \nonumber  \\
           & \stackrel{\Delta}{=}
               & \sigma_{H,\, 1}^2                 \nonumber  \\
 E
\left[
 H_{k,\, i,\, n_t,\, 1,\, I}^4
\right]    & = &  3
                 \sigma_{H,\, 1}^4                 \nonumber  \\
 E
\left[
 H_{k,\, i,\, n_t,\, 1,\, Q}^4
\right]    & = &  3
                 \sigma_{H,\, 1}^4                 \nonumber  \\
 E
\left[
 H_{k,\, i,\, n_t,\, 1,\, I}^2
 H_{k,\, j,\, n_t,\, 1,\, I}^2
\right]    & = & \sigma_{H,\, 1}^4
                 \quad\mbox{$i\ne j$}              \nonumber  \\
 E
\left[
 H_{k,\, i,\, n_t,\, 1,\, Q}^2
 H_{k,\, j,\, n_t,\, 1,\, Q}^2
\right]    & = & \sigma_{H,\, 1}^4
                 \quad\mbox{$i\ne j$}              \nonumber  \\
 E
\left[
 H_{k,\, i,\, n_t,\, 1,\, I}^2
 H_{k,\, j,\, n_t,\, 1,\, Q}^2
\right]    & = & \sigma_{H,\, 1}^4
\end{eqnarray}
       where we have used the first equation in
       (\ref{Eq:Pap10_Near_Cap_Eq3_1}) and the assumption that
       $H_{k,\, i,\, n_t,\, 1,\, I}$ and
       $H_{k,\, j,\, n_t,\, 1,\, Q}$ are independent for all $i,\, j$.
 \item Let us now compute the SNR per bit for each frame in \cite{6663392}.
       Using the notation given in \cite{6663392}, we have
\begin{eqnarray}
\label{Eq:Pap10_Near_Cap_Eq9}
\mbox{SNR}_{k,\,\mathrm{bit},\, 2}
                              \propto
                              \frac{1}{L_d}
                              \sum_{i=0}^{L_d-1}
                              \left|
                              \tilde H_{k,\, i}
                              \right|^2
\end{eqnarray}
       where the subscript $2$ in $\mbox{SNR}_{k,\,\mathrm{bit},\, 2}$
       denotes case (2). Again, the variance of the estimate of the variance
       in the right hand side of (\ref{Eq:Pap10_Near_Cap_Eq9}) is
\begin{eqnarray}
\label{Eq:Pap10_Near_Cap_Eq10}
\sigma_2^2 =
 E
\left[
\left(
\frac{1}{L_d}
\sum_{i=0}^{L_d-1}
\left|
\tilde H_{k,\, i}
\right|^2 - 2L_h\sigma^2_f
\right)^2
\right].
\end{eqnarray}
       Observe that $\tilde H_{k,\, i}$ in (\ref{Eq:Pap10_Near_Cap_Eq10}) is
       obtained by taking the $L_d$-point
       FFT of an $L_h$-tap channel, and the autocorrelation of
       $\tilde H_{k,\, i}$ is given by (37) in \cite{6663392}. Using
       Parseval's theorem we have
\begin{eqnarray}
\label{Eq:Pap10_Near_Cap_Eq11}
\frac{1}{L_d}
\sum_{i=0}^{L_d-1}
\left|
\tilde H_{k,\, i}
\right|^2 =
\sum_{n=0}^{L_h-1}
\left|
\tilde h_{k,\, n}
\right|^2
\end{eqnarray}
       where $\tilde h_{k,\, n}$ denotes a sample of zero-mean Gaussian random
       variable with variance per dimension equal to $\sigma^2_f$. Note that
       $\tilde h_{k,\, n}$ is independent over $n$ (see also (1) in
       \cite{6663392}). Substituting (\ref{Eq:Pap10_Near_Cap_Eq11}) and the
       first equation of (\ref{Eq:Pap10_Near_Cap_Eq8}) in
       (\ref{Eq:Pap10_Near_Cap_Eq10}) we get
\begin{eqnarray}
\label{Eq:Pap10_Near_Cap_Eq12}
\sigma_2^2 =
 E
\left[
\left(
\sum_{n=0}^{L_h-1}
\left|
\tilde h_{k,\, n}
\right|^2 - \sigma^2_H
\right)^2
\right].
\end{eqnarray}
       It can be shown that
\begin{eqnarray}
\label{Eq:Pap10_Near_Cap_Eq13}
\sigma_2^2 = 4L_h\sigma_f^4
\end{eqnarray}
       where we have used the following relations:
\begin{eqnarray}
\label{Eq:Pap10_Near_Cap_Eq14}
 E
\left[
\left|
\tilde h_{k,\, n}
\right|^2
\right]    & = &  2\sigma_f^2                      \nonumber \\
h_{k,\, n,\, I} +
\mathrm{j}\,
h_{k,\, n,\, Q}
           & = & \tilde h_{k,\, n}                 \nonumber  \\
 E
\left[
 h_{k,\, n,\, I}^2
\right]    & = & \sigma_f^2                        \nonumber  \\
 E
\left[
 h_{k,\, n,\, Q}^2
\right]    & = & \sigma_f^2                        \nonumber  \\
 E
\left[
 h_{k,\, n,\, I}^4
\right]    & = &  3
                 \sigma_f^4                        \nonumber  \\
 E
\left[
 h_{k,\, n,\, Q}^4
\right]    & = &  3
                 \sigma_f^4                        \nonumber  \\
 E
\left[
 h_{k,\, n,\, I}^2
 h_{k,\, m,\, I}^2
\right]    & = & \sigma_f^4
                 \quad\mbox{$n\ne m$}              \nonumber  \\
 E
\left[
 h_{k,\, n,\, Q}^2
 h_{k,\, m,\, Q}^2
\right]    & = & \sigma_f^4
                 \quad\mbox{$n\ne m$}              \nonumber  \\
 E
\left[
 h_{k,\, n,\, I}^2
 h_{k,\, m,\, Q}^2
\right]    & = & \sigma_f^4
\end{eqnarray}
       where we have assumed that $h_{k,\, n,\, I}$ and $h_{k,\, m,\, Q}$
       are independent for all $n,\, m$.
\end{enumerate}
Substituting
\begin{eqnarray}
\label{Eq:Pap10_Near_Cap_Eq15}
L_h & = & 10   \nonumber  \\
L_d & = & 4096
\end{eqnarray}
in (\ref{Eq:Pap10_Near_Cap_Eq8}) and (\ref{Eq:Pap10_Near_Cap_Eq14}) we obtain
\begin{eqnarray}
\label{Eq:Pap10_Near_Cap_Eq16}
\sigma_1^2 & = & 0.1\sigma_f^4     \nonumber  \\
\sigma_2^2 & = & 40\sigma_f^4.
\end{eqnarray}
Thus we find that the variation in the SNR per bit for each frame is 400
times larger in case (2) than in case (1). Therefore, in case (2) there are
many frames whose SNR per bit is much smaller than the average value given by
(\ref{Eq:CCT_Eq12}), resulting in a large number of bit errors. Conversely,
the average SNR per bit in case (2) needs to be much higher than in case (1)
for the same BER.

\section{Conclusions}
\label{Sec:Conclusions}
Discrete-time algorithms for the coherent detection of turbo coded MIMO
OFDM system are presented. Simulations results for a $2\times 2$
turbo coded MIMO OFDM system indicate that a BER of $10^{-5}$, is obtained
at an SNR per bit of just 5.5 dB, which is a 2.5 dB improvement over the
performance given in the literature. The minimum average SNR per bit
for error-free transmission over fading channels is derived and shown to be
equal to $-1.6$ dB, which is the same as that for the AWGN channel.

Finally, an ideal near capacity signaling is proposed, where each transmit
antenna uses a different carrier frequency. Simulation results for the
ideal coherent receiver show
that it is possible to achieve a BER of $2\times 10^{-5}$ at an average SNR
per bit equal to 2.5 dB, with two transmit and two receive antennas. When
the number of receive antennas for each transmit antenna is increased to 128,
the average SNR per bit required to attain a BER of $2\times 10^{-5}$ is
1.25 dB. The spectral efficiency of the proposed near capacity system is
1 bit/sec/Hz. Higher spectral efficiency can be obtained by increasing the
number of transmit antennas with no loss in BER performance. A pulse
shaping technique is also proposed to reduce the bandwidth of the
transmitted signal.

Future work could address the issues of peak-to-average power ratio (PAPR).

\appendix
\subsection{The Minimum Average SNR per bit for Error-free Transmission over
Fading Channels}
In this appendix, we derive the minimum average SNR per bit for error-free
transmission over MIMO fading channels. Consider the signal
\begin{eqnarray}
\label{Eq:CCT_Eq1}
\tilde r_n = \tilde x_n + \tilde w_n \qquad \mbox{for $0 \le n < N$}
\end{eqnarray}
where $\tilde x_n$ is the transmitted signal (message) and $\tilde w_n$
denotes samples of zero-mean noise, not necessarily Gaussian. All the
terms in (\ref{Eq:CCT_Eq1}) are complex-valued or two-dimensional
and are transmitted over one complex dimension.
Here the term dimension refers to a communication link between the transmitter
and the receiver carrying only real-valued signals.
We also assume that $\tilde x_n$ and $\tilde w_n$ are ergodic random processes,
that is, the time average statistics is equal to the ensemble average. The
time-averaged signal power over two-dimensions is given by, for large values of
$N$:
\begin{eqnarray}
\label{Eq:CCT_Eq2}
\frac{1}{N} \sum_{n=0}^{N-1}
\left|
\tilde x_n
\right|^2 = P_{\mathrm{av}}'.
\end{eqnarray}
The time-averaged noise power per dimension is
\begin{eqnarray}
\label{Eq:CCT_Eq3}
\frac{1}{2N} \sum_{n=0}^{N-1}
\left|
\tilde w_n
\right|^2                            =   {\sigma'}^2_w
                                     =   \frac{1}{2N}
                                         \sum_{n=0}^{N-1}
                                         \left|
                                         \tilde r_n -
                                         \tilde x_n
                                         \right|^2.
\end{eqnarray}
The received signal power over two-dimensions is
\begin{eqnarray}
\label{Eq:CCT_Eq4}
\frac{1}{N} \sum_{n=0}^{N-1}
\left|
\tilde r_n
\right|^2                          & = & \frac{1}{N}
                                         \sum_{n=0}^{N-1}
                                         \left|
                                         \tilde x_n +
                                         \tilde w_n
                                         \right|^2          \nonumber  \\
                                   & = & \frac{1}{N}
                                         \sum_{n=0}^{N-1}
                                         \left|
                                         \tilde x_n
                                         \right|^2 +
                                         \left|
                                         \tilde w_n
                                         \right|^2          \nonumber  \\
                                   & = &  P_{\mathrm{av}}' +
                                          2{\sigma'}^2_w    \nonumber  \\
                                   & = &  E
                                         \left[
                                         \left|
                                         \tilde x_n +
                                         \tilde w_n
                                         \right|^2
                                         \right]
\end{eqnarray}
where we have assumed independence between $\tilde x_n$ and $\tilde w_n$ and
the fact that $\tilde w_n$ has zero-mean. Note that in (\ref{Eq:CCT_Eq4}) it
is necessary that either $\tilde x_n$ or $\tilde w_n$ or both, have zero-mean.

Next, we observe that (\ref{Eq:CCT_Eq3}) is the expression for a
$2N$-dimensional noise hypersphere with radius $\sigma_w'\sqrt{2N}$.
Similarly, (\ref{Eq:CCT_Eq4}) is the expression for a $2N$-dimensional
received signal hypersphere with radius
$\sqrt{N(P_{\mathrm{av}}'+{2\sigma'}^2_w)}$.

Now, the problem statement is: how many noise hyperspheres (messages)
can fit into the received signal hypersphere, such that the noise
hyperspheres do not overlap (reliable decoding), for a given $N$,
$P_{\mathrm{av}}'$ and ${\sigma'}^2_w$? The solution lies in the
volume of the two hyperspheres. Note that a $2N$-dimensional hypersphere
of radius $R$ has a volume proportional to $R^{2N}$. Therefore, the number of
possible messages is
\begin{eqnarray}
\label{Eq:CCT_Eq7}
M   =   \frac{
        \left(
         N
        \left(
         P_{\mathrm{av}}'+2{\sigma'}^2_w
        \right)
        \right)^N}
             {
        \left(
         2N{\sigma'}^2_w
        \right)^N}
    =   \left(
        \frac{P_{\mathrm{av}}'+2{\sigma'}^2_w}{{2\sigma'}^2_w}
        \right)^N
\end{eqnarray}
over $N$ samples (transmissions). The number of bits required to represent
each message is $\log_2(M)$, over $N$ transmissions. Therefore, the number
of bits per transmission, defined as the channel capacity, is given by
\cite{Salehi_Dig}
\begin{eqnarray}
\label{Eq:CCT_Eq8}
C & = & \frac{1}{N} \log_2(M)                  \nonumber  \\
  & = & \log_2
        \left(
         1 +
        \frac{P_{\mathrm{av}}'}{{2\sigma'}^2_w}
        \right) \quad \mbox{bits per transmission}
\end{eqnarray}
over two dimensions or one complex dimension (here again the term
``dimension'' implies a
communication link between the transmitter and receiver, carrying only
real-valued signals. This is not to be confused with the $2N$-dimensional
hypersphere mentioned earlier or the $M$-dimensional orthogonal constellations
in \cite{Vasu_Book16}).

\begin{Propo}
\label{Propo:CC1}
Clearly, the channel capacity is additive over the number of dimensions.
In other words, channel capacity over $D$ dimensions, is equal to the
sum of the capacities over each dimension, provided the information
is independent across dimensions \cite{Vasudevan2015}.
Independence of information also implies that,
the bits transmitted over one dimension is not the interleaved version
of the bits transmitted over any other dimension.
\end{Propo}
\begin{Propo}
\label{Propo:CC2}
Conversely, if $C$ bits per transmission are sent over $2N_r$
dimensions, ($N_r$ complex dimensions), it seems reasonable to assume that
each complex dimension receives $C/N_r$ bits per transmission
\cite{Vasudevan2015}.
\end{Propo}
The reasoning for {\it Proposition\/}~\ref{Propo:CC2} is as follows. We assume
that a ``bit'' denotes ``information''. Now, if each of the $N_r$ antennas
(complex dimensions)
receive the ``same'' $C$ bits of information, then we might as well have
only one antenna, since the other antennas are not yielding
any additional information. On the other hand, if each of the $N_r$ antennas
receive ``different'' $C$ bits of information, then we end up receiving
more information ($CN_r$ bits) than what we transmit ($C$ bits), which is not
possible. Therefore, we assume that each complex dimension receives
$C/N_r$ bits of ``different'' information.

Note that, when
\begin{eqnarray}
\label{Eq:CCT_Eq9}
\tilde x_n & = & \sum_{n_t=1}^{N_t}
                 \tilde H_{k,\, n,\, n_r,\, n_t}
                  S_{k,\, 3,\, n,\, n_t}                  \nonumber  \\
\tilde w_n & = & \tilde W_{k,\, n,\, n_r}
\end{eqnarray}
as given in (\ref{Eq:Pap9_Eq37_1}), the channel capacity remains the same as 
in (\ref{Eq:CCT_Eq8}). We now define the average SNR per bit for MIMO
systems having $N_t$ transmit and $N_r$ receive antennas. We assume that
$\kappa$ information bits are transmitted simultaneously from each
transmit antenna. The amount of information received by each receive
antenna is $\kappa N_t/N_r$ bits per transmission, over two dimensions (due to
Proposition~\ref{Propo:CC2}).
Assuming independent channel frequency response and symbols across
different transmit antennas, the average SNR of
$\tilde R_{k,\, i,\, n_r}$ in (\ref{Eq:Pap9_Eq37_1}) can be computed from
(\ref{Eq:Pap9_Eq37_2}) as:
\begin{eqnarray}
\label{Eq:CCT_Eq10}
\mbox{SNR}_{\mathrm{av}}   =   \frac{2 L_h \sigma^2_f P_{\mathrm{av}} N_t}
                                    {2 L_d \sigma^2_w}
                           =   \frac{P_{\mathrm{av}}'}{{2\sigma'}^2_w}
\end{eqnarray}
for $\kappa N_t/N_r$ bits, where
\begin{eqnarray}
\label{Eq:CCT_Eq11}
P_{\mathrm{av}} =  E
                  \left[
                  \left|
                   S_{k,\, 3,\, i,\, n_t}
                  \right|^2
                  \right].
\end{eqnarray}
The average SNR per bit is
\begin{eqnarray}
\label{Eq:CCT_Eq12}
\mbox{SNR}_{\mathrm{av},\, b}
& = & \frac{2 L_h \sigma^2_f P_{\mathrm{av}} N_t}
           {2 L_d \sigma^2_w}
      \cdot
      \frac{N_r}{\kappa N_t}                               \nonumber  \\
& = & \frac{L_h \sigma^2_f P_{\mathrm{av}} N_r}
           {L_d \sigma^2_w \kappa}                         \nonumber  \\
& = & \frac{P_{\mathrm{av}}'}{{2\sigma'}^2_w}
      \cdot
      \frac{N_r}{\kappa N_t}.
\end{eqnarray}
Moreover, for each receive antenna we have
\begin{eqnarray}
\label{Eq:CCT_Eq13}
C = \kappa N_t/N_r \quad \mbox{bits per transmission}
\end{eqnarray}
over two dimensions. Substituting (\ref{Eq:CCT_Eq12}) and (\ref{Eq:CCT_Eq13})
in (\ref{Eq:CCT_Eq8}) we get
\begin{eqnarray}
\label{Eq:CCT_Eq14}
C & = & \log_2
        \left(
         1 + C \cdot \mbox{SNR}_{\mathrm{av},\, b}
        \right)                                 \nonumber  \\
\Rightarrow
\mbox{SNR}_{\mathrm{av},\, b}
  & = & \frac{2^C - 1}{C}.
\end{eqnarray}
Clearly as $C \rightarrow 0$,
$\mbox{SNR}_{\mathrm{av},\, b} \rightarrow \ln(2)$, which is the minimum
SNR required for error-free transmission over MIMO fading channels.

\bibliographystyle{IEEEtran}
\bibliography{/home/vasu/vasu/bib/mybib,/home/vasu/vasu/bib/mybib1,/home/vasu/vasu/bib/mybib2,/home/vasu/vasu/bib/mybib3,/home/vasu/vasu/bib/mybib4}
\end{document}